\documentclass[amsmath,11pt,amssymb,preprint,nofootinbib]{revtex4}
\usepackage{amsfonts} 
\usepackage{graphicx} 
\usepackage{epsfig}
\usepackage{bm}

\begin{document}

\title{Open charm mesons and Charmonia in magnetized strange hadronic matter}
\author{Amal Jahan C.S.}
\email{amaljahan@gmail.com}
\author{Amruta Mishra}
\email{amruta@physics.iitd.ac.in}
\affiliation{Department of Physics, Indian Institute of Technology,Delhi, Hauz Khas, New Delhi - 110016, India}

\begin{abstract}
We investigate the in-medium masses of open charm mesons ($D$($D^0$, $D^+$), 
$\bar{D}$($\bar{D^0}$, $D^-$), $D_s$(${D_{s}}^+$, ${D_{s}}^-$)) and 
charmonium states  ($J/\psi$, $\psi(3686)$, $\psi(3770)$, $\chi_{c0}$, 
$\chi_{c2}$) in strongly magnetized isospin asymmetric strange hadronic 
matter using a chiral effective model. In the presence of the magnetic 
field, the number density and scalar density of charged baryons have 
contributions from Landau energy levels. The mass 
modifications of open charm mesons arise due to their interactions with 
nucleons, hyperons, and the scalar fields (the non- strange field 
$\sigma$, strange field $\zeta$ and  isovector field $\delta$) 
in the presence of the magnetic field. The mass modifications 
of the charmonium states arise from the variation of dilaton 
field ($\chi$) in the magnetized medium,  which simulates the 
gluon condensates of QCD. The in-medium mass of open charm mesons 
and charmonia are observed to decrease with an increase in 
baryon density, whereas the charged $D^+$, $D^-$, ${D_{s}}^+$ 
and ${D_{s}}^-$ mesons have additional positive mass shifts 
due to Landau quantization in the presence of the magnetic field. 
The effects of strangeness fraction are found to be more dominant 
for the $\bar{D}$ mesons as compared to the $D$ mesons. 
The mass shifts of charmonia are observed to be larger 
in hyperonic medium compared to the nuclear medium. 
\end{abstract}
\maketitle
\vspace{-1.0cm}
\section{INTRODUCTION}

Strong interaction physics in the presence of large magnetic fields has gained significant attention recently. The strength of the magnetic fields in non-central  Heavy Ion Collision experiments are estimated to be $eB \sim 2{m_{\pi}}^2$ $\sim 6\times 10^{18}$ Gauss in the Relativistic Heavy Ion Collider(RHIC) at Brookhaven National Laboratory (BNL) and eB $\sim 15{m_{\pi}}^2$ $\sim 10^{19}$  Gauss in the Large Hadron Collider(LHC) at CERN \cite{kharzeev,skokov}. The electromagnetic fields are calculated using the Lienard-Wiechert potential through numerical simulations or analytical methods in these studies. The magnitude  of the magnetic field produced  is comparable to the QCD scale and this field can in turn modify the properties of hadrons at high baryon densities and/or temperature, resulting from these ultra-relativistic  high energy nuclear collisions. These medium modifications of heavy flavor mesons can affect the experimental observables such as the production and propagation of these mesons in the magnetized matter and will have phenomenological relevance. As these experiments involve nuclei which have more number of neutrons than the number of protons, the effects of isospin asymmetry are also important to be investigated. The study of magnetized hadronic matter is also important for astrophysical compact  objects like magnetars where the strength of magnetic fields may reach $10^{15}-10^{16}$ Gauss at the surface and around $10^{18}$ Gauss in the interior\cite{magnetar1,magnetar2} the discovery of novel quantum effects like chiral magnetic effect \cite{kharzeev,kharzeev_springer}, chiral separation effect \cite{kharzeev_springer,DTSon}, chiral magnetic wave \cite{chiralmagwave}, magnetic catalysis and inverse magnetic catalysis \cite{kharzeev_springer} has also motivated the study of strongly interacting matter in the presence of a magnetic field.

The properties of open heavy flavor mesons and heavy quarkonia have been investigated in QCD Sum rule approach  without accounting for the effects of magnetic field \cite{Hayashigaki,Hilger, Wang,Chabbra,Kim,Klingl,Arvindkumar_AMM_PRC_2010,Morita} and including the effects of the magnetic field\cite{Gubler,Machado,Cho,Pallabi,Rajeshkumar}. In this approach, the current-current correlation function is expanded in terms of the local operators using the operator product expansion. The QCD sum rule approach connects the spectral density of a given current correlation function through a dispersion relation with the QCD operator product expansion.
In Ref\cite{Gubler}, operators up to dimension-5 are taken into account in the operator product expansion for the vacuum part including the tensor-type quark condensate of the form $\bar{q}\sigma^{\mu\nu}q$ to investigate open charm mesons in the presence of external magnetic fields. In this study, the magnetic field induced mixing of the psuedoscalar and vector open charm mesons are found to cause significant mass modifications. The effects of Landau quantization for charged $D$ mesons are also included in this study.  In Ref \cite{Cho}, the masses of the S wave charmonium states under external magnetic fields have been studied in this QCD Sum rules approach incorporating the mixing effects between the pseudoscalar $\eta_c$ and the longitudinal component of vector $J/\psi$ state. This leads to a level repulsion in their respective mass spectra where it was observed that the mass of the  $\eta_c$ meson decreases and that of the longitudinal $J/\psi$ increases under effect of magnetic fields. The finite density calculations of the masses in nuclear matter using QCD sum rules under strong magnetic fields are incorporated in Ref \cite{Pallabi} and Ref\cite{Rajeshkumar}.

The investigation of the properties of these mesons has also been carried out using the Quark Meson Coupling (QMC) model \cite{QMC} as well as in the coupled channel approach \cite{Tolos1,Mizutani,Tolos2,Hofmann,Molina,Tolos3}. The properties of charmonium and bottomonium states have been investigated in the literature using potential models without considering magnetic fields \cite{Eichten1,Eichten2,Kluberg,Karsch,Bazavov,Digal,Mocsy,Radford} and in the presence of magnetic fields \cite{Alford,Bonati,Suzuki,Yoshida}. The most common form of potential used in these models is the Cornell potential in which the short-range interaction between the heavy quarks are governed by a Coulombic potential and the long-range interaction is governed by a linearly growing potential. The masses of heavy quarkonia are obtained by solving the Schrödinger equation using this effective potential. In the presence of the magnetic field, a spin-spin interaction potential is also included which is responsible for the mass splitting between $\eta_c$ and $J/\psi$. The singlet-triplet mixing of heavy quarkonia states is incorporated through the $-\mu.B$ term involving the spin coupling to the magnetic field \cite{Alford,Bonati,Suzuki,Yoshida}. In Ref \cite {Bonati}, the anisotropies in the static $Q\bar{Q}$ potential due to magnetic fields are taken into account which is shown to lead to an increase in the masses of heavy quarkonia. In Ref \cite{Suzuki} and Ref\cite {Yoshida}, the cylindrical Gaussian expansion method is used to obtain the anisotropic wave functions and respective energy eigenvalues for the ground state heavy quarkonia as well as the excited states. In this study, the size of the wave functions of excited states are found to be larger than that of the ground state which results in  a larger mass shift for the excited states as the magnetic field increases. 

The in-medium masses of open charm mesons, open bottom mesons, and heavy quarkonia have been studied in the literature using the chiral effective model  based on a non-linear realization of chiral symmetry and the broken scale invariance of QCD \cite{AM_PRC79_2009,AKumar_AM_EurPhys2011,Akumar_AM_PRC81_2010,Divakar_AM_IntJMod_2014,AM_Divakar_PRC_2014_2015,Divakar_AM_AdvHighEner_2015}. In this model, the underlying symmetries and the symmetry breakings of low energy QCD are realized in terms of an effective hadronic lagrangian in nuclear as well as in hyperonic medium. The model has been used  to study nuclear matter, finite nuclei \cite{Papazoglou}, hyperonic matter\cite{AM_PRC69_2004}, vector mesons  \cite{Zschiesche}, kaons and antikaons \cite{AM_PRC70_2004,AM_PRC74_2006,AM_PRC78_2008,
AM_EURPHYS_2009}, as well as to study charge-neutral matter such as the bulk matter comprising (proto)neutron stars \cite{AM_EURPHYS_2010}. Chiral symmetry is spontaneously broken in QCD which leads to non-vanishing chiral condensates and if the quarks are assumed to be massless, the scale invariance of QCD is broken by quantum effects where the non-vanishing contributions from the gluon condensate leads to the trace anomaly.  The expectation values of quark and gluon condensates in the hadronic medium  are calculated from the values of the scalar fields and the dilaton field.  The scalar fields ($\sigma$, $\zeta$, $\delta$) associated with chiral condensates along with the dilaton field $\chi$ associated with gluon condensates, interact with the heavy flavor mesons in the hadronic medium, in the chiral effective model. 

The masses of the open charm mesons are modified due to their interactions with the baryons and scalar mesons ($\sigma$, $\zeta$, $\delta$) in the medium. The medium modifications of the masses of charmonia arise due to the modifications of gluon condensates, calculated from the medium change of the dilaton field, introduced through a scale breaking term in the Lagrangian, within the chiral model. For a quarkonium bound state, if the  distance between heavy quarks $Q$ and $\bar Q$ is small compared to the characteristic scale of the non- perturbative gluonic fluctuations, then the interaction of the quarkonium state with the gluonic field can be expanded in a multipole series. In this expansion, the leading contribution to the mass of the quarkonium state is found to be a dipole term which is proportional to the gluon condensates\cite{Peskin,Voloshin}. If we assume the light quarks to be massless, then the energy-momentum tensor and hence the gluon condensate, is in fact proportional to the fourth power of dilaton field $\chi$. Hence, the mass shifts of the heavy quarkonium states arise due to the difference in the medium value of the fourth power of the dilaton field from the fourth power of its vacuum value. 

The modifications of the masses of the charmonium states in the nuclear medium have been studied in the linear density approximation using the leading order QCD formula \cite{CharmoniumLee}. In this study, the mass shift of quarkonium states in non- relativistic limit, are calculated using medium changes in the magnitude of the square of the color electric field which is called as QCD second-order Stark effect. The in-medium partial decay widths of charmonium and bottomonium states to $D\bar{D}$ pairs and $B\bar{B}$ pairs have been studied using a field theoretical model for composite hadrons \cite{AM_SPM_2015_2017} and within a light quark-antiquark pair creation model namely the $\ 3P_0$ model \cite{3P0,Friman}. In  $\ 3P_0$  model, a light quark-antiquark pair is assumed to be created in the $\ 3P_0$  state, and this light quark (antiquark) combines with the heavy charm antiquark (charm quark) of the 
decaying charmonium state, to produce the open charm $\bar{D}$ and $D$ mesons. The mass modifications of open charm mesons, open bottom mesons, charmonium states, upsilon states and the partial decay widths of charmonium and bottomonium states to $D\bar{D}$ and $B\bar{B}$ in the presence of the magnetic field in the nuclear medium within the chiral effective model have been investigated\cite{SReddy,Dhale,Amal1,AM_charmdecaywidths_mag,Amal2_upsilon,AM_SPM3,AM_SPM4,AM_SPM5,AM_SPM6}.

In this paper, we shall investigate the in-medium masses of open charm mesons ($D(D^0,D^+)$, $\bar{D}(\bar{D^0},D^-)$, $D_s ({D_S}^+,{D_S}^-)$) and charmonium states  ($J/\psi$, $\psi(3686)$, $\psi(3770)$, $\chi_{c0}$, $\chi_{c2}$) in strongly magnetized asymmetric strange hadronic  matter within the chiral effective model. The outline of the paper is as follows: In section II, we describe the chiral $\ SU(3)_L \times SU(3)_R$ effective model and its extension to  chiral SU(4) to study the open charm mesons in strange hadronic matter under an  external magnetic field. Here we include the effects of anomalous magnetic moments of the baryon octet as well as the effects of strangeness fraction in the medium. In this investigation we  do not consider the effects of temperature and  contributions from the mixing of the pseudoscalar and the vector meson states in the presence of strong magnetic fields. In section III, we present the interaction  Lagrangian of the open charm mesons with the magnetized strange hadronic matter in terms of scalar fields, number densities and scalar densities of baryons and write down the dispersion relations which  are to be solved to obtain the masses of these mesons. In  section IV, we describe the in-medium mass shifts of various charmonium states and their relation with the modification of scalar dilaton field in magnetized matter. In  section V, we discuss and analyse the results obtained and compare the results with the nuclear matter and later we summarizes our findings in section VI.

\section{THE HADRONIC CHIRAL $SU(3)_L \times  SU(3)_R$  MODEL }
In the presence of the magnetic field, the effective hadronic chiral Lagrangian density \cite{SReddy} is given as
\begin{eqnarray}
\mathcal{L_{\textrm{eff}}} = \mathcal{L_\textrm{kin}} + \sum_{W=X,Y,A,V,u}{\mathcal{L_\textrm{BW}}} + \mathcal{L_\textrm{vec}} + \mathcal{L_\textrm{0}}+ \mathcal{L_\textrm{scale break}} + \mathcal{L_\textrm{SB}} + \mathcal{L_\textrm{mag}}
\label{genlag}
\end{eqnarray}

In this equation, $\mathcal{L_\textrm{kin}}$ refers to the kinetic energy terms of the mesons and baryons. $\mathcal{L_\textrm{BW}}$ is the baryon-meson interaction term, where the index $\mathcal{\textrm{W}}$ covers both spin-0 and spin-1 mesons. Here the baryon masses are generated dynamically, through the baryon-scalar meson interactions. $\mathcal{L_\textrm{vec}}$ concerns the dynamical mass generation of the vector mesons through couplings with scalar mesons, apart from bearing the quartic self-interaction terms of these mesons. $\mathcal{L_\textrm{0}}$ contains the meson-meson interaction terms introducing the spontaneous breaking of chiral
symmetry, and $\mathcal{L_\textrm{scale break}}$ incorporates the scale invariance breaking of QCD through a logarithmic potential given in terms of scalar dilaton field $\chi$. $\mathcal{L_\textrm{SB}}$ corresponds to the explicit chiral symmetry breaking term and $\mathcal{L_\textrm{mag}}$ is the contribution by the magnetic field. We use the mean field approximation to simplify the hadronic Lagrangian density under which all the meson fields are considered as classical fields. In this approximation, only the vector fields ($\omega,\rho,\phi$) and the scalar fields (nonstrange scalar field $\sigma$, strange scalar field  $\zeta$  and scalar-isovector field $\delta$) contributes as the expectation value of the other mesons vanishes. The baryon-meson interaction term simplifies to:
\begin{eqnarray}
\mathcal{L_\textrm{BW}}=-\sum\bar\psi_{i}[ m_i^*+ g_{\omega i}\gamma_{0} \omega+  g_{\rho i}\gamma_{0} \rho+ g_{\phi i}\gamma_{0} \phi]\psi_{i}
\label{L_baryonmeson}
\end {eqnarray}
Here the index i runs over the eight lightest baryons  n, p, $\Lambda$, $\Sigma^{-}$,$\Sigma^{0}$, $\Sigma^{+}$, $\Xi^{-}$, $\Xi^{0}$ and $g_{\omega i}, g_{\rho i}, g_{\phi i}$ represent the coupling strengths of baryons with vector mesons. The effective mass of the baryons is $ m_i^* = - (g_{\sigma i}\sigma+ g_{\zeta i}\zeta + g_{\delta i}\delta$) where $g_{\sigma i}, g_{\zeta i}, g_{\delta i}$ represent the coupling strengths of baryons with scalar mesons. The other terms in the Lagrangian reduce to the following expressions,
\begin{eqnarray}
\mathcal{L_\textrm{vec}}=
\frac{1}{2}( m_\omega^2\omega^2 + m_\rho^2\rho^2+ m_\phi^2\phi^2)(\frac{\chi^2}{{\chi_0}^2})+g_4(\omega^4+6\rho^2\omega^2+\rho^4+2\phi^4)
\label{Lvec}
\end{eqnarray}
\begin{eqnarray}
\mathcal{L_\textrm{0}}+\mathcal{L_\textrm{scale break}}=-\frac{1}{2}k_0\chi^2(\sigma^2+\zeta^2+\delta^2)+k_1(\sigma^2+\zeta^2+\delta^2)^2\nonumber\\+k_2(\frac{\sigma^4}{2}+\frac{\delta^4}{2}+3\sigma^2\delta^2+\zeta^4)+k_3\chi(\sigma^2-\delta^2)\zeta\nonumber\\-k_4\chi^4 -\chi^4\ln{(\frac{\chi}{\chi_0})}+\frac{d}{3}\chi^4\ln\bigg(\big(\frac{(\sigma^2-\delta^2)\zeta}{{\sigma_0^2}\zeta_0}\big)\big(\frac{\chi}{\chi_0}\big)^3\bigg)
\label{L0_scale}
\end{eqnarray}
\begin{eqnarray}
\mathcal{L_\textrm{SB}}=-\Big(\frac{\chi}{\chi_0}\Big)^2 \Big[{m_\pi}^2f_\pi \sigma +(\sqrt{2}{m_K}^2f_K-\frac{1}{\sqrt{2}}{m_\pi}^2f_\pi)\zeta\Big]
\label{L_SB}
\end{eqnarray}Finally the contribution of magnetic field incorporated in the Lagrangian term given by
\begin{eqnarray}
\mathcal{L_\textrm{mag}} =-\bar{\psi_{i}}q_{i}\gamma_{\mu}A^{\mu}\psi_{i}-\frac{1}{4}\kappa_{i}\mu_{N}\bar{\psi_{i}}\sigma^{\mu\nu}F_{\mu\nu}\psi_{i}-\frac{1}{4}F^{\mu\nu}F_{\mu\nu}
\label{Lmag}
\end{eqnarray}
The second term in eq.(\ref{Lmag}) which is a tensorial interaction term  is related to the anomalous magnetic moment(AMM) of the baryons. In this term,  $\mu_{N}$ is the nuclear Bohr magneton, given as $\mu_{N}$  = $e/(2m_N)$, where $m_N$ is the vacuum mass of the nucleon. $\kappa_{i}$ is the gyromagnetic ratio corresponding to the anomalous magnetic moment of the baryons and the values $\kappa_{i}$ used in our calculations were taken from Refs \cite{Broderick1_Rabhietal} and \cite{Wei}. We choose the magnetic field to be uniform and along the z axis, and take the vector potential to be $A^{\mu}$ = (0, 0, Bx, 0). From the mean field Lagrangian density, the coupled equations of motion for the scalar fields $\sigma$, $\zeta$, $\delta$, $\chi$ and vector meson fields $\omega, \rho, \phi$ are obtained in terms of scalar densities and number densities of baryons \cite{AM_PRC69_2004,Zschiesche}. The magnetic fields introduce summation over Landau levels in the expressions of number density and scalar density of charged baryons ( i= p, $\Sigma^{-}$, $\Sigma^{+}$, $\Xi^{-}$ ) which are given as \cite {Wei,Mao,Broderick2}
\begin{equation}
\rho_i = \frac{eB}{2\pi^2}\Bigg[{\sum_{\nu}^{{\nu^{(S=1)}_{max}}} k_{f, \nu, 1}^i} + {\sum_{\nu}^{{\nu^{(S=-1)}_{max}}} k_{f, \nu, -1}^i}\Bigg]
\label{charged_numberdensity}
\end{equation}
\begin{eqnarray}
\rho_s^i = \frac{eBm_i^*}{2\pi^2}\Bigg[{\sum_{\nu}^{\nu^{(S=1)}_{max}} \frac{\sqrt[]{m_i^{*2} + 2eB\nu} + \Delta_i}{\sqrt[]{m_i^{*2} + 2eB\nu}} ln\Bigg|\frac{k_{f,\nu,1}^i + E_f^i}{\sqrt[]{m_i^{*2} + 2eB\nu} + \Delta_i}\Bigg|} \nonumber \\
+ {\sum_{\nu}^{\nu^{(S=-1)}_{max}} \frac{\sqrt[]{m_i^{*2} + 2eB\nu} - \Delta_i}{\sqrt[]{m_i^{*2} + 2eB\nu}} \ln\Bigg|\frac{k_{f,\nu,-1}^i + E_f^i}{\sqrt[]{m_i^{*2} + 2eB\nu} - \Delta_i}\Bigg|}\Bigg]
\label{charged_scalardensity}
\end{eqnarray}
Here $k_{f, \nu, s}^i$ is the Fermi momentum of charged baryons, $ E_f^i$ is the Fermi energy, $\nu$ is the Landau level and spin index S = +1($-1$) corresponds to spin up (spin down) projections  for the baryons. The parameter $\Delta_{i}$  refers to the anomalous magnetic moments of the baryons given as $\Delta_{i}=-\frac{1}{2}\kappa_i\mu_N B$. The Landau levels of charged baryons is enumerated using the expression  $\nu = n+\frac{1}{2}-\frac{q_B}{|q_B|}\frac{S}{2}$ where $q_B$ is the charge of the baryon($q_B=e$ for p, $\Sigma^{+}$ and $q_B=-e$ for $\Sigma^{-}$, $\Xi^{-}$). The lowest Landau level for a particular spin projection of the charged baryon is obtained by setting n=0 in this expression. The maximum allowed value of Landau level for a charged baryon is determined using the expression $ 
 \nu_{max} = \Big\lfloor\frac{(E_f^i - S\Delta_i)^2 - m_i^{*2}}{2eB}\Big\rfloor$ where the floor operator acting on a quantity x, ie. $\lfloor x\rfloor$ is defined as the largest integer less than or equal to x. The Fermi momenta of charged baryons are related to their Fermi energies $ E_f^i$ as 
\begin{eqnarray}
k_{f, \nu, S}^i = \sqrt[]{(E_f^i)^2 - \Big(\sqrt[]{m_i^{*2} + 2eB\nu} + s\Delta_i\Big)^2}
\label{charged_fermimomentum}
\end{eqnarray}
For neutral baryons ( i= n,$\Lambda$, $\Sigma^{0}$, $\Xi^{0}$ ), there is no Landau quantization contribution in the presence of an external magnetic field. The number density and scalar density are given as

\begin{eqnarray}
\rho_i = \frac{1}{4\pi^2}\sum_{S=\pm1} \Bigg(\frac{2}{3}(k_{f,S}^i)^3 + S\Delta_i\Bigg[(m_i^* + S\Delta_i)k_{f, S}^i + \nonumber \\ (E_f^i)^2\Bigg\{arcsin\Bigg(\frac{m_i^* + S\Delta_i}{E_f^i}\Bigg) - \frac{\pi}{2}\Bigg\}\Bigg]\Bigg) 
\label{neutral_numberdensity}
\end{eqnarray}
\begin{eqnarray}
\rho_s^i = \frac{m_i^*}{4\pi^2}\sum_{S=\pm1}\Bigg[k_{f,S}^iE_f^i - (m_i^* + S\Delta_i)^2\ln\Bigg|\frac{k_{f,S}^i + E_f^i}{m_i^* + S\Delta_i}\Bigg|\Bigg]
\label{neutral_scalardensity}
\end{eqnarray}
The Fermi momenta of neutral baryons $k_{f,S}^i$ are related to their Fermi energies $ E_f^i$ as 
\begin{eqnarray}
k_{f, S}^i = \sqrt[]{(E_f^i)^2 - (m_i^* + S\Delta_i)^2}
\label{neutral_fermimomenta}
\end{eqnarray}

We consider the hyperonic matter in equilibrium which results in five chemical equilibrium equations of baryons which are given as
\begin{eqnarray}
\Lambda + \Lambda\rightleftharpoons p+{\Xi^-}\\ 
\Lambda + \Lambda\rightleftharpoons n+\Xi^0\\ 
\Sigma^-+p \rightleftharpoons \Lambda + n\\
\Sigma^++n \rightleftharpoons \Lambda+p\\
\Sigma^0+\Sigma^-\rightleftharpoons \Xi^-+n\hspace{-0.5cm}
\label{Hyperoneqbm}
\end{eqnarray}
The above equations constrain the chemical potentials $\mu_i$ of all the baryons in the medium. Since the particles on left and right hand sides are in chemical equilibrium, their chemical potentials can be equated and given as
\begin{eqnarray}
2\mu_\Lambda=\mu_p+\mu_{\Xi^-}\hspace{0.50cm}\\  2\mu_\Lambda=\mu_n+\mu_{\Xi^0}\hspace{0.50cm}\\  \mu_{\Sigma^-}+\mu_p=\mu_n+\mu_{\Lambda}\\  \mu_{\Sigma^+}+\mu_n=\mu_p+\mu_{\Lambda}\\ \mu_{\Sigma^-}+\mu_{\Sigma^0}=\mu_n+\mu_{\Xi^-} \hspace{-0.4cm}
\label{chempot_hyperebm}
\end{eqnarray}

These equations can be further rewritten in terms of the effective chemical potential of baryons $\mu_i^*$ through the relation $\mu_i^* =\mu_i-(g_{\rho i}\rho+g_{\omega i}\omega+g_{\phi i}\phi)$. The effective chemical potential $\mu_i^*$ is numerically equal to the Fermi energy of baryons at temperature T=0. These chemical equilibrium equations provide the necessary  equations of constraint in studying the strange hadronic matter under external magnetic fields. The equations of motion of scalar fields are then solved self consistently at different magnetic fields for given values of total baryon density $\rho_B= \sum_i\rho_i$, isospin asymmetry parameter $\eta= \frac{-\sum_i{I_{3i}\rho_i}}{\rho_B}$ and strangeness fraction, $f_s= \frac{\sum_i|{S_i|\rho_i}}{\rho_B}$ where $I_{3i}$ is the third component of isospin and  $S_i$ is the strangeness quantum number for the $i^{th}$ baryon. The strangeness fraction is a measure of the relative population of hyperons (with appropriate weight factors due to the number of strange quarks in the hyperons) among all the baryons present in the medium.

In the following section, we shall describe the interaction  of open charm mesons with strongly magnetized  strange hadronic matter and the medium  modifications of their masses.

\section{IN-MEDIUM MASSES OF THE OPEN CHARM MESONS }
We examine the medium modifications for the open charm meson masses in asymmetric  magnetized strange hadronic matter. These mesons interact with light quark condensates, which are modified significantly in the hadronic medium. Here  the chiral
SU(3) has been generalized to chiral SU(4) to include the charmed mesons and their interactions  with the light hadronic
sector \cite{AM_PRC79_2009, AKumar_AM_EurPhys2011, Akumar_AM_PRC81_2010,AM_PRC70_2004}.

The interaction Lagrangian density of $D$ and $\bar{D}$ mesons with the strange hadronic medium in chiral effective model is given as\cite{AM_PRC78_2008}

\begin{eqnarray}
{\cal L}_{int} & = & -\frac {i}{8 f_D^2} 
\Big [3\Big (\bar p \gamma^\mu p
+\bar n \gamma ^\mu n \Big) 
\Big(\Big({D^0} (\partial_\mu \bar D^0)
- (\partial_\mu {{D^0}}) {\bar D}^0 \Big )
+\Big(D^+ (\partial_\mu D^-) - (\partial_\mu {D^+})  D^- \Big )\Big )
\nonumber \\
& +&
\Big (\bar p \gamma^\mu p -\bar n \gamma ^\mu n \Big) 
\Big( \Big({D^0} (\partial_\mu \bar D^0) - (\partial_\mu {{D^0}}) 
{\bar D}^0 \Big )
- \Big( D^+ (\partial_\mu D^-) - (\partial_\mu {D^+})  D^- \Big )\Big )
\nonumber\\
&+& 2\Big((\bar{\Lambda}^{0}\gamma^{\mu}\Lambda^{0})
 \Big( \Big({D^0} (\partial_\mu \bar D^0)
 -(\partial_\mu {{D^0}}) {\bar D}^0 \Big)
+ \Big(D^+ (\partial_\mu D^-) - (\partial_\mu D^+)  D^- \Big) \Big)\nonumber\\
 &+& 2 \Big(\Big(\bar{\Sigma}^{+}\gamma^{\mu}\Sigma^{+}
 + \bar{\Sigma}^{-}\gamma^{\mu}\Sigma^{-}\Big)
 \Big(\Big({D^0} (\partial_\mu \bar D^0)
-(\partial_\mu {{D^0}}) {\bar D}^0 \Big)
+ \Big(D^+ (\partial_\mu D^-) - (\partial_\mu D^+)  D^- \Big)\Big)\nonumber\\
&+& \Big(\bar{\Sigma}^{+}\gamma^{\mu}\Sigma^{+}
 - \bar{\Sigma}^{-}\gamma^{\mu}\Sigma^{-}\Big)
 \Big(\Big({D^0} (\partial_\mu \bar D^0)
 -(\partial_\mu {{D^0}}) {\bar D}^0 \Big)
- \Big(D^+ (\partial_\mu D^-) - (\partial_\mu D^+)  
D^- \Big)\Big)\Big)\nonumber\\
&+&2\Big(\bar{\Sigma}^{0}\gamma^{\mu}\Sigma^{0}\Big)
 \Big(\Big({D^0} (\partial_\mu \bar D^0)
 -(\partial_\mu {{D^0}}) {\bar D}^0 \Big)
+ \Big(D^+ (\partial_\mu D^-) - (\partial_\mu D^+)  D^- \Big) \Big)\nonumber\\
 &+& \Big(\bar{\Xi}^{0}\gamma^{\mu}\Xi^{0}
 + \bar{\Xi}^{-}\gamma^{\mu}\Xi^{-}\Big)
 \Big(\Big({D^0} (\partial_\mu \bar D^0)
 -(\partial_\mu {{D^0}}) {\bar D}^0 \Big)
+ \Big(D^+ (\partial_\mu D^-) - (\partial_\mu D^+)  D^- \Big)\Big)\nonumber\\
&+&\Big(\bar{\Xi}^{0}\gamma^{\mu}\Xi^{0}
 - \bar{\Xi}^{-}\gamma^{\mu}\Xi^{-}\Big)
 \Big(\Big({D^0} (\partial_\mu \bar D^0)
 -(\partial_\mu {{D^0}}) {\bar D}^0 \Big)
- \Big(D^+ (\partial_\mu D^-) - (\partial_\mu D^+)  
D^- \Big)\Big)\Big ]\nonumber\\
&+& \frac{m_D^2}{2f_D} \Big [ 
(\sigma +\sqrt 2 \zeta_c)\big (\bar D^0 { D^0}+(D^- D^+) \big )
+\delta \big (\bar D^0 { D^0})-(D^- D^+) \big )
\Big ] \nonumber \\
& - & \frac {1}{f_D}\Big [ 
(\sigma +\sqrt 2 \zeta_c )
\Big ((\partial _\mu {{\bar D}^0})(\partial ^\mu {D^0})
+(\partial _\mu {D^-})(\partial ^\mu {D^+}) \Big )\nonumber\\
  &+& \delta
\Big ((\partial _\mu {{\bar D}^0})(\partial ^\mu {D^0})
-(\partial _\mu {D^-})(\partial ^\mu {D^+}) \Big )
\Big ]\nonumber \\
&+& \frac {d_1}{2 f_D^2}(\bar p p +\bar n n +\bar{\Lambda}^{0}\Lambda^{0}+\bar{\Sigma}^{+}\Sigma^{+}+\bar{\Sigma}^{0}\Sigma^{0}
+\bar{\Sigma}^{-}\Sigma^{-}+\bar{\Xi}^{0}\Xi^{0}+\bar{\Xi}^{-}\Xi^{-})
  \big ( (\partial _\mu {D^-})(\partial ^\mu {D^+})\nonumber \\
&+&(\partial _\mu {{\bar D}^0})(\partial ^\mu {D^0})
\big )
+ \frac {d_2}{2 f_D^2} \Big [
\Big(\bar p p+\frac{1}{6}\bar{\Lambda}^{0}\Lambda^{0}
+\bar{\Sigma}^{+}\Sigma^{+}+\frac{1}{2}\bar{\Sigma}^{0}\Sigma^{0}\Big)
(\partial_\mu {\bar D}^0)(\partial^\mu {D^0})\nonumber \\
&+&\Big(\bar n n+\frac{1}{6}\bar{\Lambda}^{0}\Lambda^{0}
+\bar{\Sigma}^{-}\Sigma^{-}+\frac{1}{2}\bar{\Sigma}^{0}\Sigma^{0}\Big)
 (\partial_\mu D^-)(\partial^\mu D^+)\Big ]
\label{lddbar}
\end{eqnarray}

The interaction Lagrangian density of $D_s$ mesons with the strange hadronic medium is given as\cite{Divakar_AM_AdvHighEner_2015}
\begin{equation}
\begin{array}{l}
{\cal L}_{int}=-\frac{i}{4 f_{D_{S}}^{2}}\big[(2(\bar{\Xi}^{0} \gamma^{\mu} \Xi^{0}+\bar{\Xi}^{-} \gamma^{\mu} \Xi^{-})+\bar{\Lambda}^{0} \gamma^{\mu} \Lambda^{0}+\bar{\Sigma}^{+} \gamma^{\mu} \Sigma^{+} +\bar{\Sigma}^{0} \gamma^{\mu} \Sigma^{0}+\bar{\Sigma}^{-} \gamma^{\mu} \Sigma^{-})\\(D_{S}^{+}(\partial_{\mu} D_{S}^{-})-(\partial_{\mu} D_{S}^{+}) D_{S}^{-})\big] 
+\frac{m_{D_{S}}^{2}}{\sqrt{2} f_{D_{S}}}\big[(\zeta^{\prime}+\zeta_{c}^{\prime})(D_{S}^{+} D_{S}^{-})\big] 
-\frac{\sqrt{2}}{f_{D_{S}}}\big[(\zeta^{\prime}+\zeta_{c}^{\prime})((\partial_{\mu} D_{S}^{+})(\partial^{\mu} D_{S}^{-}))\big] \\
+\frac{d_{1}}{2 f_{D_{S}}^{2}}\big[(\bar{p} p+\bar{n} n+\bar{\Lambda}^{0} \Lambda^{0}+\bar{\Sigma}^{+} \Sigma^{+}+\bar{\Sigma}^{0} \Sigma^{0}+\bar{\Sigma}^{-} \Sigma^{-}+\bar{\Xi}^{0} \Xi^{0}+\bar{\Xi}^{-} \Xi^{-})((\partial_{\mu} D_{S}^{+})(\partial^{\mu} D_{S}^{-}))\big] \\
+\frac{d_{2}}{2 f_{D_{S}}^{2}}\big[(2(\bar{\Xi}^{0} \Xi^{0}+\bar{\Xi}^{-} \Xi^{-})+\bar{\Lambda}^{0} \Lambda^{0}+\bar{\Sigma}^{+} \Sigma^{+}+\bar{\Sigma}^{0} \Sigma^{0}+\bar{\Sigma}^{-} \Sigma^{-})((\partial_{\mu} D_{S}^{+})(\partial^{\mu} D_{S}^{-}))\big]
\end{array}
\label{LDs}
\end{equation}

In eq.(\ref{lddbar}) and (\ref{LDs}), the first term is the vectorial Weinberg-Tomozawa interaction term, obtained from the kinetic energy term of eq.(\ref{genlag}). The second term, which is the scalar meson exchange term, is obtained from the explicit symmetry-breaking
term. The next three terms of above Lagrangian density are known as the range terms. The first range term  is obtained from the kinetic energy term of the pseudoscalar mesons where $f_D$ and $f_{D_s}$ refers to the decay constants of $D$ and $D_s$ mesons respectively. The parameters $d_1$ and $d_2$ in the last two range terms are determined by a fit of the empirical values of the Kaon-Nucleon scattering lengths\cite{Brown,Bielich,Barnes_PRC49} for
I = 0 and I = 1 channels\cite{AM_PRC78_2008, AM_EURPHYS_2009}. The interaction Lagrangian density gives rise to equations of motion for $D$, $\bar{D}$ and $D_s$ mesons and their Fourier transforms lead to the dispersion relations given as
\begin{eqnarray}
-\omega^2 + \overrightarrow{k}^2 + m_{j}^2 - \Pi_{j}(\omega,|\overrightarrow{k}|) = 0
\label{disp_relation}
\end{eqnarray}

Here the index j denotes the various open charm mesons $D$, $\bar{D}$ and $D_s$ and $ m_{j}$ is the vacuum mass of the corresponding open charm meson  and $\Pi_{j}(\omega,|\vec{k}|)$ denotes the self energy of the open charm mesons in the medium. For $D$ mesons the self energy is given as
\begin{eqnarray}
 \Pi_{{D}} (\omega, |\vec k|) &= & \frac {1}{4 f_D^2}\Big [3 (\rho_p +\rho_n)
\pm (\rho_p -\rho_n) 
+2\big(\left( \rho_{\Sigma^{+}}+ \rho_{\Sigma^{-}}\right) \pm
\left(\rho_{\Sigma^{+}}- \rho_{\Sigma^{-}}\right) \big)\nonumber\\
&+&2(\rho_{\Lambda^{0}}+\rho_{\Sigma^{0}})
+( \left( \rho_{\Xi^{0}}+ \rho_{\Xi^{-}}\right) 
\pm 
\left(\rho_{\Xi^{0}}- \rho_{\Xi^{-}}\right)) 
\Big ] \omega \nonumber \\
&+&\frac {m_D^2}{2 f_D} (\sigma ' +\sqrt 2 {\zeta_c} ' \pm \delta ')
+ \Big [- \frac {1}{f_D}
(\sigma ' +\sqrt 2 {\zeta_c} ' \pm \delta ')
+\frac {d_1}{2 f_D ^2} (\rho_p ^s +\rho_n ^s\nonumber\\
&+&\rho_{\Lambda^{0}}^s+\rho_{\Sigma^{+}}^s+\rho_{\Sigma^{0}}^s
+\rho_{\Sigma^{-}}^s
+\rho_{\Xi^{0}}^s+\rho_{\Xi^{-}}^s)
+\frac {d_2}{4 f_D ^2} \Big ((\rho _p^s +\rho_n^s)
\pm   ({\rho} _p^s -{\rho}_n^s)+\frac{1}{3}{\rho} _{\Lambda^0}^s\nonumber\\
&+&({\rho}_{\Sigma^{+}}^s +{\rho} _{\Sigma^{-}}^s)
\pm    ({\rho} _{\Sigma^{+}}^s -{\rho}_{\Sigma^{-}}^s)
+{\rho} _{\Sigma^{0}}^s \Big ) \Big ]
(\omega ^2 - {\vec k}^2)
\label{selfenergy_D}
\end{eqnarray}
where $\pm$ refers to $D^0$ and $D^+$ respectively. For $\bar{D}$ mesons the self energy is given as
\begin{eqnarray}
 \Pi_{\bar{D}}(\omega, |\vec k|) &= & -\frac {1}{4 f_D^2}\Big [3 (\rho_p +\rho_n)
\pm (\rho_p -\rho_n)
 +2\big(\left( \rho_{\Sigma^{+}}+ \rho_{\Sigma^{-}}\right)\pm 
\left(\rho_{\Sigma^{+}}- \rho_{\Sigma^{-}}\right) \big)\nonumber\\
&+&2(\rho_{\Lambda^{0}}+\rho_{\Sigma^{0}})
+( \left( \rho_{\Xi^{0}}+ \rho_{\Xi^{-}}\right) 
\pm 
\left(\rho_{\Xi^{0}}- \rho_{\Xi^{-}}\right))\Big ] \omega\nonumber \\
&+&\frac {m_D^2}{2 f_D} (\sigma ' +\sqrt 2 {\zeta_c} ' \pm \delta ')
 + \Big [- \frac {1}{f_D}
(\sigma ' +\sqrt 2 {\zeta_c} ' \pm \delta ')
+\frac {d_1}{2 f_D ^2} ({\rho}_p^s +{\rho}_n^s\nonumber\\
&+&{\rho}_{\Lambda^{0}}^s+{\rho}_{\Sigma^{+}}^s
+{\rho}_{\Sigma^{0}}^s+{\rho}_{\Sigma^{-}}^s
+{\rho}_{\Xi^{0}}^s+{\rho}_{\Xi^{-}}^s)
+\frac {d_2}{4 f_D ^2} \Big (({\rho}_p^s +{\rho}_n^s)
\pm   ({\rho}_p^s -{\rho}_n^s)+\frac{1}{3}{\rho}_{\Lambda^{0}}^s
\nonumber\\
&+&({\rho} _{\Sigma^{+}}^s +{\rho} _{\Sigma^{-}}^s)
\pm ({\rho}_{\Sigma^{+}}^s -{\rho}_{\Sigma^{-}}^s)
+{\rho} _{\Sigma^{0}}^s \Big ) \Big ]
(\omega ^2 - {\vec k}^2)
\label{selfenergy_Dbar}
\end{eqnarray}
where $\pm$ refers to $\bar{D^0}$ and $D^-$ respectively. The expression for self energy for strange-charmed mesons reads:
\begin{equation}
\begin{array}{l}
\Pi_{D_s}(\omega,|\vec{k}|)=\left[\left(\frac{d_{1}}{2 f_{D_{S}}^{2}}\left(\rho_{p}^{s}+\rho_{n}^{s}+\rho_{\Lambda}^{s}+\rho_{\Sigma^{+}}^{s}+\rho_{\Sigma^{0}}^{s}+\rho_{\Sigma^{-}}^{s}+\rho_{\Xi^{0}}^{s}+\rho_{\Xi^{-}}^{s}\right)\right)\right.\\
+\left(\frac{d_{2}}{2 f_{D_{S}}^{2}}\left(2\left(\rho_{\Xi^{0}}^{s}+\rho_{\Xi^{-}}^{s}\right)+\rho_{\Lambda}^{s}+\rho_{\Sigma^{+}}^{s}+\rho_{\Sigma^{0}}^{s}+\rho_{\Sigma^{-}}^{s}\right)\right)\\
\left.-\left(\frac{\sqrt{2}}{f_{D_{S}}}\left(\zeta^{\prime}+\zeta_{c}^{\prime}\right)\right)\right]\left(\omega^{2}-\vec{k}^{2}\right) \\
 \pm\left[\frac{1}{2 f_{D_{S}}^{2}}\left(2\left(\rho_{\Xi^{0}}+\rho_{\Xi^{-}}\right)+\rho_{\Lambda}+\rho_{\Sigma^{+}}+\rho_{\Sigma^{0}}+\rho_{\Sigma^{-}}\right)\right] \omega 
+\left[\frac{m_{D_{S}}^{2}}{\sqrt{2} f_{D_{S}}}\left(\zeta^{\prime}+\zeta_{c}^{\prime}\right)\right]
\end{array}
\label{selfenergy_Ds}
\end{equation}

Here $\pm$ signs in the co-efficient of $\omega$ refers to ${D_S}^+$ and ${D_S}^-$ mesons respectively.
In eq.(\ref{selfenergy_D}), eq.(\ref{selfenergy_Dbar}) and eq. (\ref{selfenergy_Ds}), $\sigma^\prime$ = ($\sigma - \sigma_0$), $\zeta^\prime_c$ = ($\zeta_c - \zeta_{c0}$), $\delta^\prime$ = ($\delta - \delta_0$) and $\zeta^\prime$ = ($\zeta - \zeta_0$). The fluctuation  $\zeta^\prime_c$ has been observed to be negligible \cite{Roder} and its
contribution to the in-medium masses of open charm mesons will be neglected in the present investigation. The charged open charm mesons ( j= $D^+$, $D^-$, ${D_S}^+$, ${D_S}^-$) have additional positive mass modification in magnetic fields which, retaining only the lowest Landau level,
is given as
\begin{eqnarray}
m^{eff}_{j} = \sqrt[]{m_{j}^{*2} + |eB|}
\label{charged_effective_mass}
\end{eqnarray}
where $m_{j}^{*}$ are solutions for $\omega$ at $|\overrightarrow{k}|$ = 0 of the dispersion relations given by eq.(\ref{disp_relation}). For the neutral open charm mesons ( j= $D^0$, $\bar{D^0}$), there is no contribution from the landau quantization effects and the effective mass in the medium are given as, 
\begin{eqnarray}
m^{eff}_{j} = {m_{j}}^{*} 
\label{neutral_effective_mass}
\end{eqnarray}
\section{MASS SHIFTS OF CHARMONIUM STATES}
In this section we shall be dealing with the in- medium masses of charmonia  $(c\bar{c})$ such as $J/\psi$, $\psi(3686)$, $\psi(3770)$, $\chi_{c0}$, $\chi_{c2}$, which are 1S, 2S, 1D, $1^3P_0$ and $1^3P_2$ states respectively in strange hadronic matter in the presence of strong magnetic fields. The heavy quarkonium states are modified in a hadronic environment due to modifications of gluon condensates\cite{Peskin,Voloshin,CharmoniumLee}. The trace anomaly in QCD indicates that the trace of energy momentum tensor in QCD is non zero when scale symmetry is broken. A non-zero trace of the energy-momentum tensor in QCD is known to originate from the gluon condensates and finite quark mass contributions. This scale invariance breaking is simulated in the chiral SU(3) model using the scale breaking Lagrangian (last two terms in eq.(\ref{L0_scale}) involving a scalar, gluon, color-singlet dilaton field $\chi$. Comparing the expressions for the trace of energy-momentum tensor from the trace anomaly and from the scale breaking Lagrangian after neglecting finite quark mass contributions we get\cite{Cohen,Heide}.
\begin{equation}
\Theta_{\mu}^{\mu}=\left\langle\frac{\beta_{Q C D}}{2 g} G_{\mu \nu}^{a} G^{\mu \nu a}\right\rangle \equiv-(1-d) \chi^{4}
\label{energymom_tensor}
\end{equation} where the parameter d originates from the second logarithmic term scale breaking Lagrangian and one loop QCD $\beta$ function is given as
\begin{equation}\beta_{\mathrm{QCD}}(g)=-\frac{11 N_{c} g^{3}}{48 \pi^{2}}\left(1-\frac{2 N_{f}}{11 N_{c}}\right)
\label{QCD_Beta}
\end{equation}
Here, $N_c=3$ refers to the number of colors and $N_f$ denotes the number of quark flavors. In the above equation, the first term in the parentheses arises from the antiscreening contribution of the gluons and the second term arises from the screening
contribution of quark pairs.

If we neglect  the masses  of  the quarks then the in-medium masses of charmonia are calculated using modifications of gluon condensates simulated by the  medium  change  of  the  dilaton  field $\chi$, under  an  external  magnetic  field. In the Chiral Effective model the leading order mass shift formula of the charmonium states in the large charm mass limit is given as\cite{AKumar_AM_EurPhys2011,Akumar_AM_PRC81_2010, Amal1}
\begin{equation}
\Delta m_{\psi}= \frac{4}{81} (1 - d) \int dk^{2} 
\langle \vert \frac{\partial \psi (\vec k)}{\partial {\vec k}} 
\vert^{2} \rangle
\frac{k}{k^{2} / m_{c} + \epsilon}  \left( \chi^{4} - {\chi_0}^{4}\right), 
\label{masspsi}
\end{equation}
where 
\begin{equation}
\langle \vert \frac{\partial \psi (\vec k)}{\partial {\vec k}} 
\vert^{2} \rangle
=\frac {1}{4\pi}\int 
\vert \frac{\partial \psi (\vec k)}{\partial {\vec k}} \vert^{2}
d\Omega
\label{integralsymbol}
\end{equation}

 In the above, $\ m_{c}$ is the mass of the corresponding heavy quark and $\epsilon$ = $2m_{c}- m_{\Psi}$ represents the binding energy of the corresponding charmonium state and $\chi$ and $\chi_0 $ are the values of the dilaton field in the magnetized medium and in the vacuum respectively. $\psi(k)$ is the wave function in the momentum space normalized as $\int \frac{d^{3} k}{(2 \pi)^{3}}|\psi(k)|^{2}=1$ . The wave functions for these heavy quarkonium states are taken to be harmonic oscillator wave functions and are given as\cite{Friman}
\begin{eqnarray}
\psi_{N,l} =N {Y_{l}}^m ( \theta, \phi)(\beta ^2r^2)^{l/2}   e^{-\frac{1}{2} \beta ^2r^2} L_{N-1}^{l+\frac{1}{2}} (\beta ^2r^2)
\label{wavefunction}
\end{eqnarray}
where $\beta^2$=M$\omega$/h characterizes the strength of the harmonic potential with $M = m_{c}/2 $  and ${L_{p}}^{k}(z)$ is the associated Laguerre Polynomial. The $\beta$ values for $J/\psi$, $\psi(3686)$, $\psi(3770)$ are obtained by fitting their root mean squared radii\cite{CharmoniumLee} and those of $\chi_{c0}$ and $\chi_{c2}$ are subsequently obtained by the linear extrapolation of the $\beta$ vs vacuum mass of the charmonium states\cite{AM_charmdecaywidths_mag}.

\section{RESULTS AND DISCUSSION}
In this section, we discuss the numerical results of the modifications of the scalar fields and subsequently the masses of open charm mesons and the mass shifts charmonium states in the isospin strange hadronic matter under strong magnetic fields. From the chiral effective Lagrangian (eq.(\ref{genlag})), the equations of motion of scalar fields and vector fields are obtained and are solved self consistently. The values of parameters in the chiral effective model are chosen to be $g_{\sigma N}$ = 10.6 and  $g_{\zeta N}$ = $-0.47$ which are determined by fitting to the baryon masses in the vacuum. The other parameters calculated by fitting to the  saturation properties of asymmetric nuclear matter in the mean field theory are: $g_{\omega N}$ = 13.3, $g_{\rho p}$ = 5.5, $g_4$ = 79.7, $g_{\delta p}$ = 2.5, $m_\zeta$ = 1024.5 MeV, $m_\sigma$ = 466.5 MeV and $m_\delta$ = 899.5 MeV. The values of the couplings of hyperons with the scalar fields are $g_{\sigma\Lambda}$ = 7.52, $g_{\zeta\Lambda}$ = 5.8, $g_{\delta\Lambda}$ = 0, $g_{\sigma\Sigma}$ = 6.13, $g_{\zeta\Sigma}$ = 5.8, $g_{\delta{\Sigma}^+}$ = 6.79, $g_{\delta{\Sigma}^-} = -6.79 $, 
$g_{\delta{\Sigma}^0}$ = 0,    $g_{\sigma\Xi}$ = 3.78, $g_{\zeta\Xi}$ = 9.14, $g_{\delta{\Xi}^0}$ = 2.36 and
$g_{\delta{\Xi}^-}= -2.36$. The values of couplings of hyperons with the vector mesons are $g_{\omega\Lambda}$ = $g_{\omega\Sigma}$ = $\frac{2}{3}$ $g_{\omega N}$, $g_{\rho{\Sigma}^+}$ = $\frac{2}{3}$ $g_{\omega N}$,
$g_{\rho{\Sigma}^-}$= $-\frac{2}{3}$ $g_{\omega N}$, $g_{\rho{\Sigma}^0}$ = 0, $g_{\omega\Xi}$ =$\frac{1}{3}$ $g_{\omega N}$, $g_{\rho\Lambda}$ = 0, $g_{\rho{\Xi}^0}$ = $\frac{1}{3}$ $g_{\omega N}$, $g_{\rho{\Xi}^-}$ = $-\frac{1}{3}$ $g_{\omega N}$, $g_{\phi\Lambda}$=$g_{\phi\Sigma}$=$-\frac{\sqrt{2}}{3}$$g_{\omega N}$, $g_{\phi\Xi}$=$-\frac{2\sqrt{2}}{3}$ $g_{\omega N}$. The values of other parameters used in our calculation are $k_0$=2.54, $k_1$=1.35, $k_2=-4.78$, $k_3 =-2.77$ and $k_4=-0.22$ and d=0.064. The vacuum values of scalar fields denoted as $\sigma_0$, $\zeta_0$, and $\chi_0$  are $-93.3$ MeV, $-106.6$ MeV and 409.8 MeV respectively.

\begin{figure}[htbp]
\includegraphics[height=17.6cm, width=16.0cm]{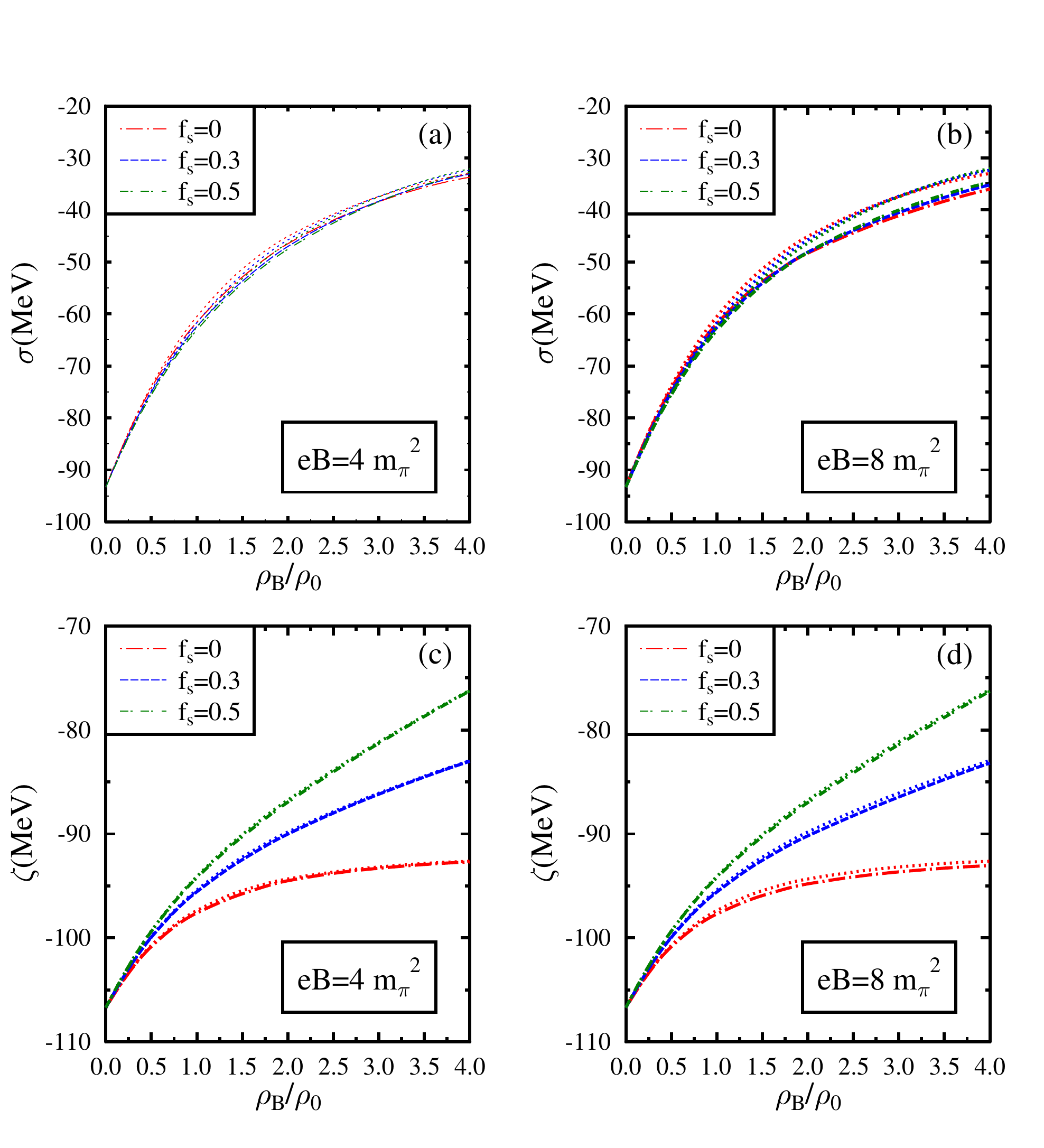}
\caption{The scalar fields $\sigma$ and $\zeta$ plotted as function of baryon density for both nuclear ($f_s$ = 0), and hyperonic ($f_s$ = 0.3, 0.5) matter situations at magnetic fields $eB=4m_{\pi}^2$ as well as for $eB=8m_{\pi}^2$, for isospin asymmetry parameter ($\eta=0.5$) , when the effects of anomalous magnetic moment are taken into account (dashed lines), and compared to the case when the effects of anomalous magnetic moment are not taken into account (dotted line)}
\label{sigma_zeta}
\end{figure}

\begin{figure}[htbp]
\includegraphics[height=17.6cm, width=16.0cm]{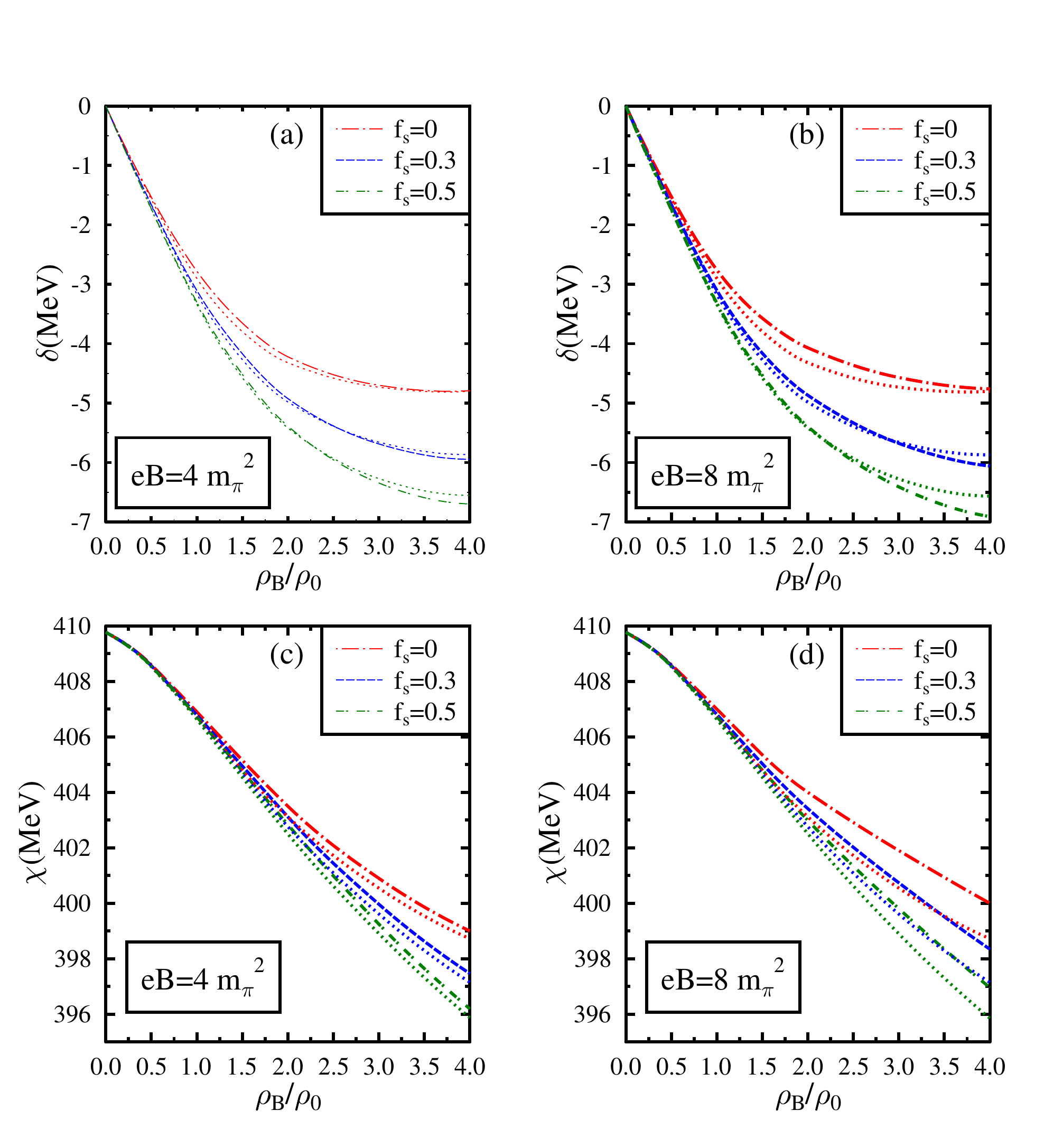}
\caption{The scalar fields $\delta$ and $\chi$ plotted as function of baryon density for both nuclear ($f_s$ = 0), and hyperonic ($f_s$ = 0.3, 0.5) matter situations at magnetic fields $eB=4m_{\pi}^2$ as well as for $eB=8m_{\pi}^2$, for isospin asymmetry parameter ($\eta=0.5$) , when the effects of anomalous magnetic moment are taken into account (dashed lines), and compared to the case when the effects of anomalous magnetic moment are not taken into account (dotted line)}
\label{delta_chi}
\end{figure}

In Figures \ref{sigma_zeta} and \ref{delta_chi}, the values of the scalar fields $\sigma$, $\zeta$, $\delta$  and $\chi$  are plotted as functions of baryon density $\rho_{B} $/$\rho_{0} $ (where $\rho_{0} $ is nuclear matter saturation density)  at magnetic fields  eB= $4{m_{\pi}}^2$ and $8{m_{\pi}}^2$ for isospin asymmetry parameter $\eta$=0.5 and strangeness fraction $f_{s}$=0 (corresponding to pure nuclear matter) as well as $f_{s}$= 0.3, 0.5 (corresponding to strange hadronic matter). The effects from the anomalous magnetic moments (AMMs) of the baryons are also included in these plots and are compared with the case where the effects from the anomalous magnetic moments are ignored (shown as dotted lines). The magnitude of the scalar fields $\sigma$ and $\zeta$, are observed to decrease as the baryon density increases in the isospin asymmetric strange hadronic matter ($\eta=0.5$) under a constant magnetic field. This is due to the increase of scalar density of the baryons with an increase in baryon density. The strange scalar field $\zeta$ at high densities show a saturation behavior, which is observed in the $f_{s}$=0 situation. This effect becomes less pronounced in the hyperonic matter where the variation of $\zeta$ as a function of density is more pronounced since the coupling of $\zeta$  with hyperons is larger than it's coupling with nucleons. Since the equation of motion of $\chi$ is coupled with that of $\sigma$, $\zeta$ and $\delta$ through the scalar meson interaction and the scale breaking Lagrangian terms, the  magnitude of dilaton field $\chi$ also drops with an increase in density. In contrast to the above, the magnitude of $\delta$ initially increases with density and then tends to a saturation behavior at high densities. For nuclear matter this saturation behaviour of $\delta$ is observed at densities above $\rho_{B}$=2.5 $\rho_{0}$.

At lower densities and at a fixed magnetic field, as strangeness fraction $f_s$ increases from 0 to 0.5, the magnitude of $\sigma$  increases. But as we go to high densities, the magnitude of $\sigma$ instead decreases as  $f_s$  increases. This behavioral change in the dependence of $\sigma$ on  $f_s$  happens at approximately $\rho_B$=3.3 $\rho_0$ for $eB=4m_{\pi}^2$ and at approximately $\rho_B$=1.9 $\rho_0$ for $eB=8m_{\pi}^2$ when AMM effects are considered. When AMM effects are not taken into account, the change in the dependence of $\sigma$ on $f_s$ happens at approximately $\rho_B$=3.2 $\rho_0$ for $eB=4m_{\pi}^2$ as well as for $eB=8m_{\pi}^2$. This is because, at high densities, the sum of the scalar densities of baryons in the strange hadronic medium is larger than the scalar density of neutrons in the pure nuclear matter in the $\eta=0.5$ situation. We observe that the magnitude of $\zeta$  decreases as strangeness fraction increases especially at large densities. The magnitude of $\delta$ is observed to increase with increase in $f_s$ and the onset of saturation behavior of this field is observed at higher densities in the hyperonic matter compared to the pure nuclear matter situation. The dilaton field $\chi$ is observed to decrease as $f_s$ increases from 0 to 0.5 due to its coupling with the other scalar fields. The effect of $f_s$ on $\sigma$ and $\delta$ are prominent when magnetic fields are stronger.

The scalar densities and number densities of the baryons are modified with the consideration of anomalous magnetic moment (AMM) effects in the presence of the magnetic field due to the tensorial interaction term. In the expressions for scalar densities of the baryons, the effects of the magnetic field are in terms of the summation over the Landau energy levels (for the charged baryons) and the anomalous magnetic moments of the baryons. In the absence of AMM, the neutral baryons are not subjected to the effect of the magnetic field. The effect of the anomalous magnetic moments is observed to give slightly smaller values of the scalar densities of most of the baryons as compared to the case when these effects are neglected. The scalar density of nucleons are especially more affected by AMM effects compared to hyperons in the presence of the magnetic field as the values of the gyromagnetic ratio corresponding to the anomalous magnetic moments of the nucleons are slightly larger\cite{Broderick1_Rabhietal}. This leads to the magnitude of $\sigma$ with AMM effects being larger. The effects of AMM are seen to be larger at high densities and larger magnetic fields. In the case of the $\zeta$ field, the effect of AMM is similar to that of $\sigma$ but the variation is extremely small in magnitude. The dependence of the magnitude of $\delta$ on AMM effects are elevated at high densities, at large strangeness fraction and, at stronger magnetic fields. It is observed that at large densities the magnitude of $\delta$ increases when AMM effects are taken into account.

The effects of the magnetic field on scalar fields are observed to be  less dominant compared to the effects of density. For a given strangeness fraction and density, at $\eta=0.5$, the magnitude of the scalar fields $\sigma$ and $\chi$ increases marginally with increase in the magnetic field when AMM effects are taken into account. At baryon density $\rho_B=4\rho_0$ there is much more appreciable increase  of ($\sim$ 2 MeV) in the magnitude of $\sigma$ and ($\sim$ 1 MeV) in the magnitude of $\chi$ as the magnetic field increases from eB= $\ 4m_\pi^2$ to eB= $\ 8m_\pi^2$ when AMM effects are included. The variation of $\zeta$ as function of the magnetic field is observed to be even smaller in the isospin asymmetric hyperonic matter compared to other scalar fields. The value of $\delta$ is almost constant as a function of the magnetic field at lower densities but at higher densities (above 3$\rho_0$ ) it is observed to increase with increase in the magnetic field when AMM effects are taken into account for strange hadronic matter. For $\eta$=0.5 when AMM effects are neglected, the effect of magnetic field on scalar fields are extremely small in the hyperonic matter situation compared to the case where the AMM effects are included. In this case only the charged baryons among the baryon octet will be affected due to the variation in the magnetic field.

 For isospin asymmetry parameter $\eta=0.5$, strangeness fraction  $f_s=0.3$ and for external magnetic field eB= $\ 4m_\pi^2$, the values of $\sigma$ in MeV  are observed to be $-62.42$ $(-61.68)$, $-47.02$ $(-45.77)$, $-38.31$ $(-37.37)$  and $-33.08$($-32.36$) at $\rho_B$= 1$\rho_0$, 2$\rho_0$, 3$\rho_0$ and 4$\rho_0$ respectively with (without) AMM effects. For eB= $\ 8m_\pi^2$ they are modified to $ -62.57$ $(-61.66)$,$-48.11$ $(-45.76)$,$-40.49$ $(-37.35)$ and $-35.17$ $(-32.35)$ respectively. For eB= $\ 4m_\pi^2$ the values of $\zeta$ in MeV  are observed to be $-95.55$ $(-95.40)$, $-90.01$ $(-89.82)$, $-86.17$ $(-86.08)$ and  $-83.01(-82.96)$ and for eB= $\ 8m_\pi^2$ they become $-95.58$ $(-95.39)$,$-90.18$($-89.81$),
$-86.44$ $(-86.06)$ and 
$-83.18$ $(-82.94)$  at $\rho_B$= 1$\rho_0$, 2$\rho_0$, 3$\rho_0$ and 4$\rho_0$ respectively with (without) AMM effects. In the case of $\delta$, the values in same order are $ -3.10$ $(-3.16)$,$-4.92$ $(-4.97)$, $-5.68$ $(-5.65)$ and $-5.94$ $(-5.86)$ for  eB= $\ 4m_\pi^2$ and as the magnetic field is increased to  eB= $\ 8m_\pi^2$, the values of  $\delta$ are modified to $-3.10$ $(-3.16)$, $-4.86$ $(-4.98)$,$-5.67$ $(-5.66)$ and $-6.05$  $(-5.87)$. The values of dilaton field $\chi$ under the same medium conditions are found to be 406.78 (406.65),403.11 (402.76), 399.96 (399.61), 397.46 (397.14) for eB= $\ 4m_\pi^2$ and 406.80 (406.65),403.41 (402.75), 400.73 (399.60), 398.34 (397.14) for eB= $\ 8m_\pi^2$  at $\rho_B$= 1$\rho_0$, 2$\rho_0$, 3$\rho_0$ and 4$\rho_0$
respectively with (without) AMM effects.

\begin{figure}[htbp]
\includegraphics[height=17.6cm, width=16.0cm]{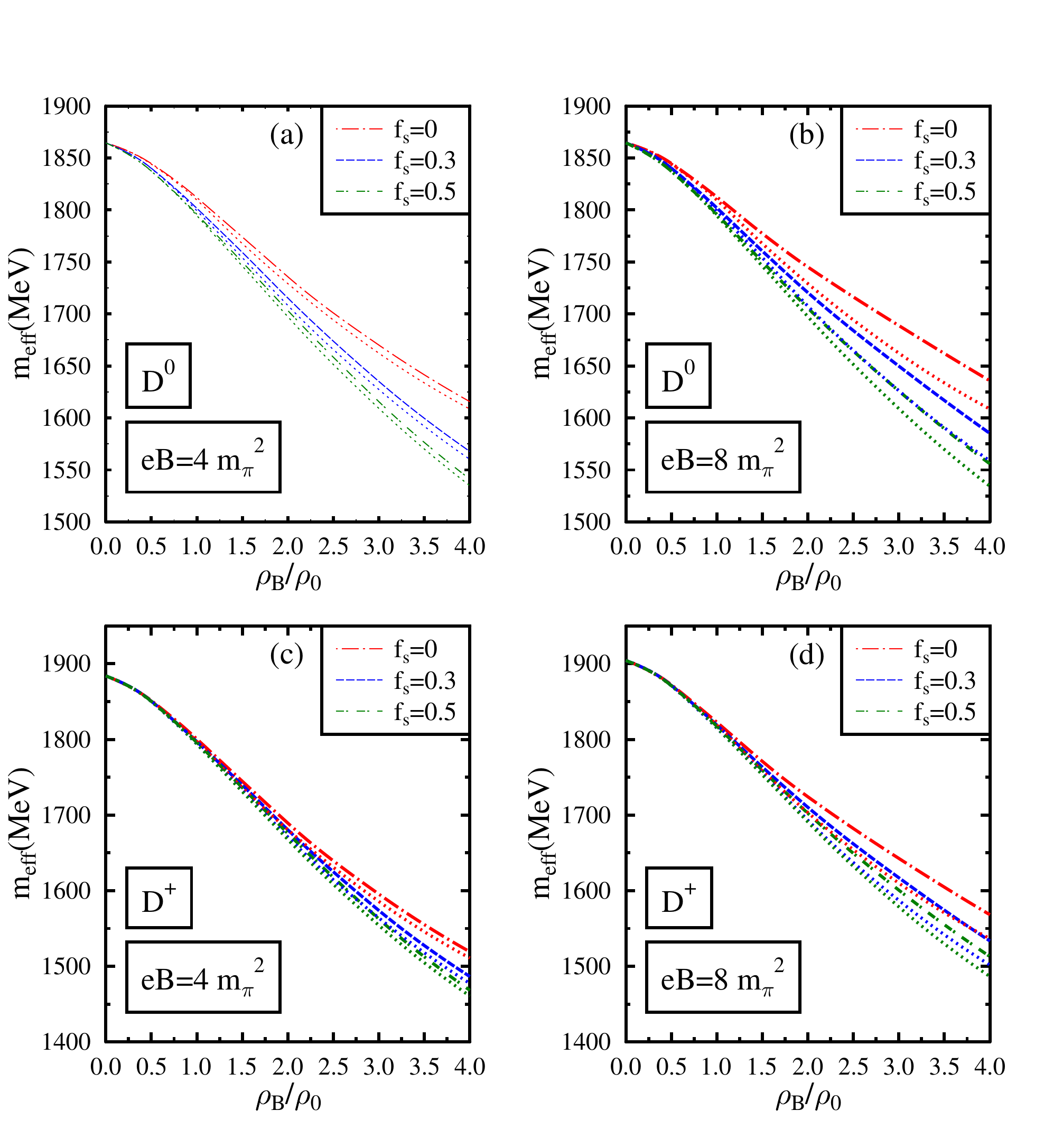}
\caption{The effective masses of $D(D^0,D^+)$ mesons in MeV, plotted as function of baryon density for both nuclear ($f_s$ = 0), and hyperonic ($f_s$ = 0.3, 0.5) matter situations at magnetic fields $eB=4m_{\pi}^2$ as well as for $eB=8m_{\pi}^2$, for isospin asymmetry parameter ($\eta=0.5$) , when the effects of anomalous magnetic moment are taken into account (dashed lines), and compared to the case when the effects of anomalous magnetic moment are not taken into account (dotted line)}
\label{mD}
\end{figure}

\begin{figure}[htbp]
\includegraphics[height=17.6cm, width=16.0cm]{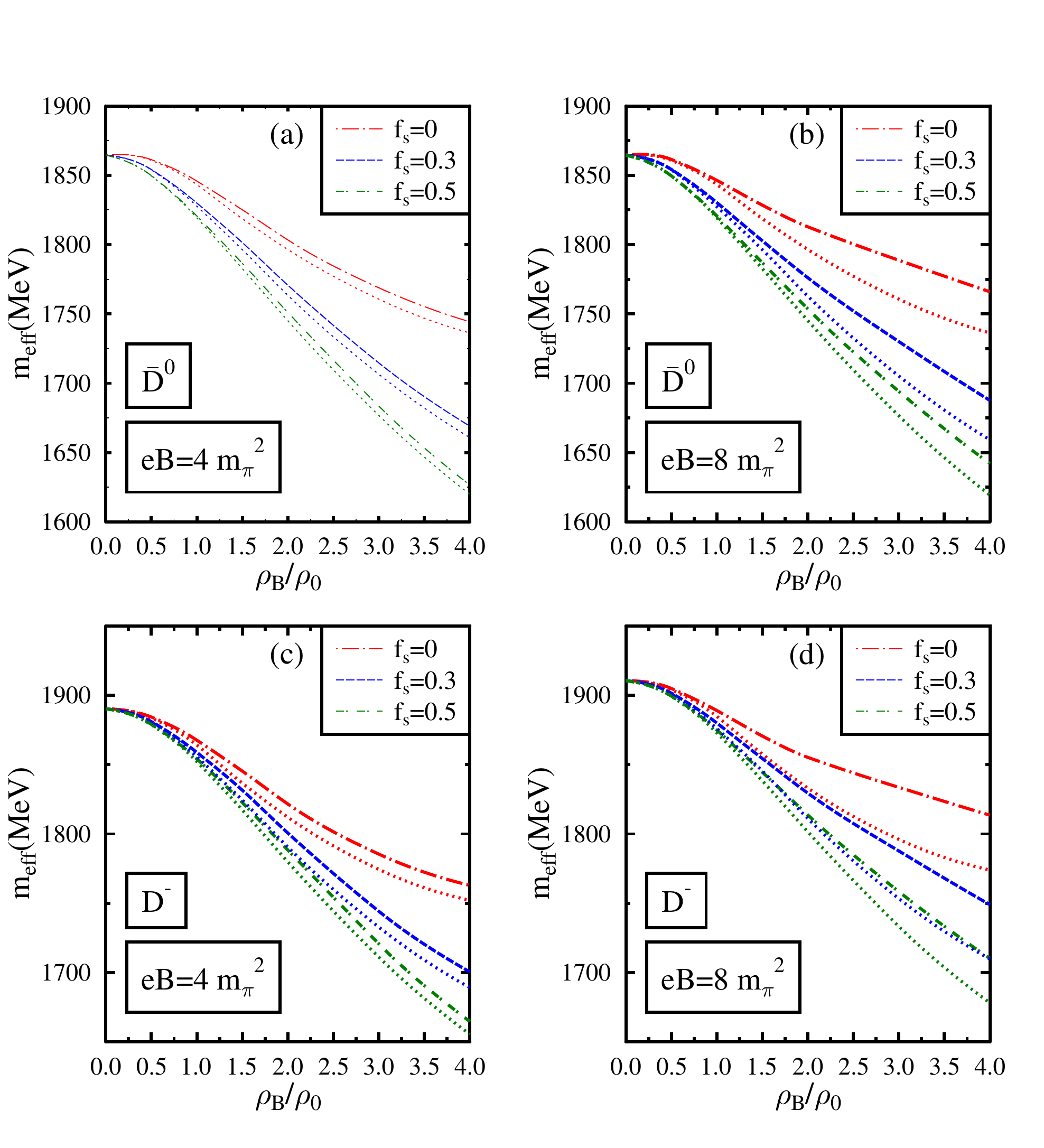}
\caption{The effective masses of $\bar{D}(\bar{D^0},D^-)$  mesons in MeV, plotted as function of baryon density for both nuclear ($f_s$ = 0), and hyperonic ($f_s$ = 0.3, 0.5) matter situations at magnetic fields $eB=4m_{\pi}^2$ as well as for $eB=8m_{\pi}^2$, for isospin asymmetry parameter ($\eta=0.5$) , when the effects of anomalous magnetic moment are taken into account (dashed lines), and compared to the case when the effects of anomalous magnetic moment are not taken into account (dotted line)}
\label{mDbar}
\end{figure}

\begin{figure}[htbp]
\includegraphics[height=17.6cm, width=16.0cm]{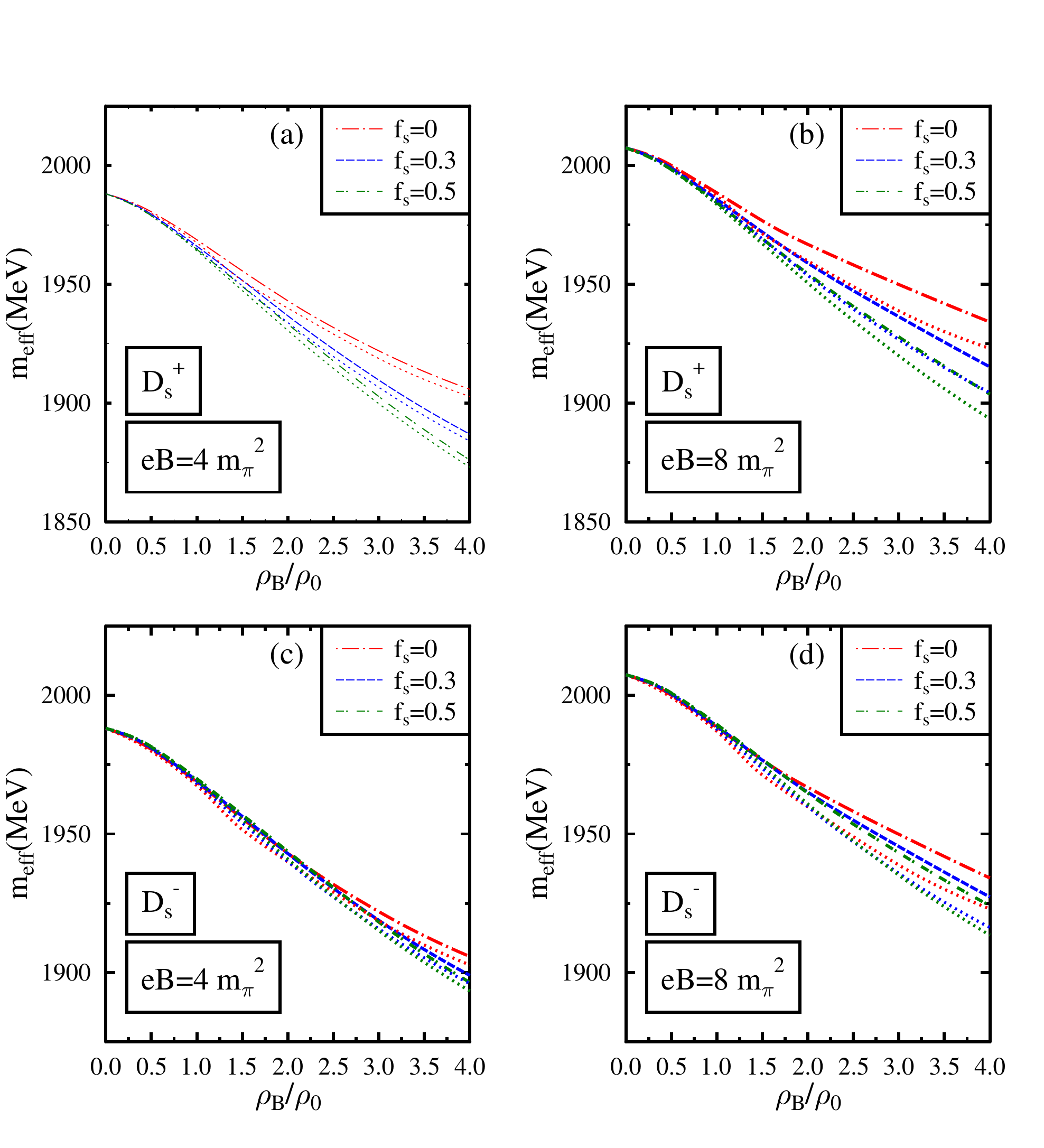}
\caption{The effective masses of $({D_s}^+,{D_s}^-)$ mesons in MeV, plotted as function of baryon density for both nuclear ($f_s$ = 0), and hyperonic ($f_s$ = 0.3, 0.5) matter situations at magnetic fields $eB=4m_{\pi}^2$ as well as for $eB=8m_{\pi}^2$, for isospin asymmetry parameter ($\eta=0.5$) , when the effects of anomalous magnetic moment are taken into account (dashed lines), and compared to the case when the effects of anomalous magnetic moment are not taken into account (dotted line)}
\label{mD_s}
\end{figure}

The in-medium masses of the open charm mesons $D(D^0,D^+)$, $\bar{D}(\bar{D^0},D^-)$ and $D_s({D_s}^+,{D_s}^-)$ in the presence of magnetic fields   are calculated using the dispersion relations for these mesons and plotted in figures \ref{mD}, \ref{mDbar} and \ref{mD_s} respectively. The values of parameters used in our calculations are  $d_1$= 2.56/$m_K$ and $d_2$ =0.73/$m_K$. The D meson decay constant will be taken to be $f_D$=135 MeV \cite{Morath} and their vacuum masses are $m_{D^+}$ = $m_{D^-}$ = 1869 MeV and $m_{D^0}$ = $m_{\bar{D^0}}$ = 1864.5 MeV. In case of the $D_s$ meson, the decay constant is taken to be $f_{D_s}$= 235 MeV \cite{Divakar_AM_AdvHighEner_2015} and their vacuum masses are $m_{{D_s}^+}$ = $m_{{D_s}^-}$ = 1968.5 MeV. The modifications in the masses of these mesons are due to the Weinberg-Tomozawa term, the scalar-exchange term arising from the explicit symmetry-breaking term, the range terms and from the direct contribution of the magnetic field through Landau quantization effects for charged mesons in accord with eq.(\ref{charged_effective_mass}).

The isospin symmetric part of Weinberg-Tomozawa term is attractive for $D$ mesons (cause a drop of the masses of $D^0$ and $D^+$ ) and repulsive for $\bar{D}$ mesons (leads to an increase in the mass of $\bar{D^0}$ and $D^-$). The Weinberg-Tomozawa term results in the mass splitting of  mesons in the isospin doublets in the asymmetric hadronic medium giving a positive contribution to the mass of $D^0$ and negative contribution to the mass of  $D^+$. Similarly, when isospin asymmetry is introduced, $D^-$ mass is observed to have an increase whereas the mass of $\bar{D^0}$ drops. This is because the Weinberg-Tomozawa term  has asymmetric contributions dependent on the number densities of baryons which distinguish between these isospin pairs. Hence $D^+$ and $D^0$  as well as $D^-$ and $\bar{D^0}$ mesons have non-degenerate masses in the asymmetric matter. One must note that in the presence of the magnetic field, even in the symmetric matter, the mass degeneracy is broken by the non zero value of $\delta$ due to the Landau quantization effects of charged baryons. This non zero value of $\delta$  contributes to the scalar meson exchange term and range terms which result in a mass splitting in the symmetric matter.  

 In the asymmetric nuclear matter($\eta=0.5$) since $\delta$ is negative, the $\delta$ term part of the scalar-meson exchange term gives a drop in the masses for charged  $D^+$ and $D^-$ mesons and gives positive contribution to $D^0$ and $\bar{D^0}$ mesons compared to the symmetric case. The $\delta$ term makes the first range term more repulsive for charged  $D^+$ and $D^-$ mesons and less repulsive for $D^0$ and $\bar{D^0}$ mesons in the asymmetric nuclear matter. The $d_1$ range term is attractive for all mesons but the $d_2$ term cause a drop in the mass of charged $D^+$ and $D^-$ mesons and does not contribute to neutral $D^0$ and $\bar{D^0}$ mesons for $\eta=0.5$ nuclear matter. The medium modifications of $D$ and $\bar{D}$ mesons in the magnetized nuclear medium due to the contribution of these terms are already discussed in Ref\cite {SReddy}.

 The in medium masses of the $D$ and $\bar{D}$ shows a drop from their vacuum masses as the baryon density increases in the strange hadronic medium for a fixed value of the magnetic field and strangeness fraction. This is due to the attractive  $d_1$ and $d_2$ range terms which dominate over the repulsive first range term at large densities along with the attractive scalar meson exchange term. The $d_1$ and $d_2$ terms contain scalar densities of baryons which increase with density and the scalar interaction terms contain the fluctuations of scalar fields $\sigma$ and $\delta$ which also grows with density. The attractive contributions of the above terms at higher densities also dominate over the repulsive contributions of Weinberg- Tomozawa term for $\bar{D}$ mesons. In the case of $D$ mesons, the Weinberg-Tomozawa is itself attractive in nature. Hence the masses of $D$ ($D^0$, $D^+$) mesons drop more than
$\bar{D}$(($\bar{D^0}$, $D^-$)) mesons as baryon density increases, Due to cumulative isospin asymmetric contributions of other terms, among $D^0$ and $D^+$ mesons, the latter drops more as density increases. At $f_s$=0.3, under a magnetic field of eB= $\ 4m_\pi^2$, incorporating the effects of AMM, $D^0$ and $D^+$ experiences a mass drop of approximately 63 MeV and 72 MeV from their vacuum values at $\rho_B$= 1$\rho_0$ where as $\bar{D^0}$ and $D^-$ experiences a mass drop of 35 MeV and 11 MeV respectively. At $\rho_B$= 4$\rho_0$ the mass drops of $D^0$ and  $D^+$ increases to  297 MeV  and 383 MeV respectively and the mass drops of $\bar{D^0}$ and $D^-$ becomes 205 MeV and 168 MeV.

At $\eta=0.5$, for a given value of the magnetic field and density, as we move from the nuclear matter to the hyperonic matter, $D$ and $\bar{D}$ experience a larger mass drop.
For instance, under a magnetic field of eB= $\ 4m_\pi^2$, at $\rho_B$= 1$\rho_0$. incorporating the effects of AMM, the mass drops of $D^0$ and $D^+$ mesons are 10 Mev and 3 MeV more in in $f_s$=0.3 case when compared with the nuclear matter. Under the same conditions, the mass drops of $\bar{D^0}$ and $D^-$ mesons are 16 MeV and 9 MeV more in $f_s$=0.3 compared to $f_s$=0.0 situation. When the baryon density increases to $\rho_B$= 4$\rho_0$, $D^0$, $D^+$, $\bar{D^0}$ and $D^-$ mesons have an additional mass drop of 48 MeV, 33 MeV, 58 MeV  and 62 MeV  respectively in $f_s$=0.3 compared to the pure nuclear matter situation. The above behaviour is due to the fact that the sum of baryonic scalar densities increases with an increase in strangeness fraction especially at large densities. The $d_1$ and  $d_2$  range terms depend on the baryonic scalar densities. As strangeness fraction increases, the  dominant $d_1$ term becomes more attractive compared to the $d_2$ term. Hence the cumulative attractive contribution of total range term which increases with an increase of $f_s$ in the medium ensures large drops in the meson masses relative to the nuclear matter situation. Moreover, the contribution of the  Weinberg-Tomozawa term to the in-medium mass of the $D$ and $\bar{D}$ mesons becomes weaker when the medium is populated with hyperons \cite{AKumar_AM_EurPhys2011}. Hence it further contributes to the decrease in the mass of $\bar{D}$ mesons in the strange hadronic medium as the repulsive contributions becomes weaker. In the case of $D$ mesons, the attractive contributions of Weinberg- Tomozawa term become weaker in the presence of hyperons. Thus $\bar{D}$ mesons are subjected to larger mass drop than the $D$ mesons with an increase in the strangeness fraction of the medium.

The asymmetric contributions in the Weinberg-Tomozawa term and $d_2$ range term are dominantly responsible for the difference in the mass drop of isospin doublets with strangeness in the medium. The attractive scalar meson
exchange term as well as the first range term is reliant on the fluctuations of the scalar fields  $\sigma$ and $\delta$ and the former does not behave in a monotonic manner with an increase in the value of $f_s$ as discussed before. But the magnitude of  $\delta$ increases with an increase in strangeness fraction at all densities. Hence the magnitude of the scalar meson exchange and first range term depends on the interplay of fluctuation of both these fields in the strange hadronic medium. They also contributes to the difference in the mass drop of charged $D$ mesons in comparison with their neutral partners.

The presence of the magnetic field introduces modifications of $D$ and $\bar{D}$ mesons by modifying the values of the scalar fields and the scalar densities of baryons. For a fixed value of density and strangeness fraction, the mass of $D$ and $\bar{D}$ mesons are observed to increase with an increase in the magnetic field when AMM effects are taken into consideration especially at high densities. This is because for isosopin asymmetric matter, as the magnetic field increases the cumulative scalar density of the baryons decreases and the magnitude of $\sigma$ increases. This reduces the magnitude of the attractive contributions from the $d_1$ and $d_2$ range terms which depends on the scalar densities as well as the scalar meson exchange term which depends on the fluctuation of $\sigma$. The neutral $D$ mesons are subjected to less change compared to  the charged $D$ mesons as a function of the magnetic field since the latter provide additional positive mass modifications for $D^+$ and $D^-$ mesons due to Landau quantization in accord with eq.(\ref{charged_effective_mass}). The $D^0$ and $\bar{D^0}$  mesons shows only a marginal modification in its mass with respect to change in the magnetic field at nuclear saturation
density. For $f_s=0.3$, incorporating the effects of AMM, as the magnetic field is increased from eB= $\ 4m_\pi^2$ to eB= $\ 8m_\pi^2$, $D^0$ and $\bar{D^0}$ experience an increase of mass of  approximately 17 MeV and 18 MeV respectively at $\rho_B$= 4$\rho_0$ whereas $D^+$  and $D^-$ experience an increase of 47 MeV and 49 MeV  respectively in their in-medium mass.

The repulsive contribution of the magnetic field on the in-medium masses of $D$ and $\bar{D}$ mesons is slightly larger in the pure nuclear medium compared to the hyperonic medium for the isospin asymmetry parameter $\eta=0.5$ and $f_s=0.5$. This is because, in the nuclear matter at a fixed baryon density, the magnitude of the scalar field $\delta$ exhibits a marginal drop with an increase in the magnetic field. But in the hyperonic matter at $f_s=0.5$, the magnitude of $\delta$ has an increasing trend with respect to the increase in the magnetic field. The magnitude of $\sigma$ is observed to increase with an increase in the magnetic field for both nuclear and hyperonic matter. The amount of increase in the magnitude of $\sigma$ with respect to the magnetic field at a fixed baryon density is less in the hyperonic matter compared to the nuclear matter situation. Hence both the fluctuations in $\sigma$ and $\delta$ from their respective vacuum values is less in magnitude in nuclear medium as the magnetic field increases. This has a larger effect on the attractive scalar meson exchange term and total range term in such a manner that the overall magnitude of these attractive terms weakens more in the nuclear medium compared to the hyperonic matter with respect to the increase in the magnetic field. This results in the overall contributions of repulsive terms slightly stronger in the nuclear medium as a function of magnetic field.

When the anomalous magnetic moments(AMM) of the baryons are ignored, only the charged baryons in the baryon octet are subjected to change in the magnetic field. For isospin asymmetry parameter $\eta=0.5$ and strangeness fraction $f_s=0.0$, the nuclear matter is composed of only neutrons and when AMM effects are not taken into consideration, neutrons will not undergo modifications in the magnetic field as they are electrically neutral. But for non zero values of strangeness fraction, the medium is populated by hyperons and due to the presence of the charged particles among them, their scalar densities are subjected to change in the magnetic field. But the magnitude of variation of scalar densities and scalar fields with the change in the magnetic field is small when AMM effects are ignored and the variation increases with an increase in $f_s$. The in-medium mass of $D$ and $\bar{D}$ mesons are observed to be smaller when AMM effects are not taken into account compared to the cases where these effects are included. This is due to the larger value of scalar densities and smaller magnitude of $\sigma$ when AMM effects are not present which increases the attractive contributions from the total range term and scalar meson exchange term. The effect of magnetic field on $D^0$ and $\bar{D^0}$ are extremely minute in this case such that they experience a drop of less than 1-2 MeV in their mass as the magnetic field increases from eB =$\ 4m_\pi^2$ to eB =$\ 8m_\pi^2$.

The in-medium masses of $D_s$ mesons drop as baryon density increases in the hadronic medium when the value of the magnetic field is fixed. In the case of the nuclear medium, there is no contribution from the Weinberg-Tomozawa term as well as from the $d_2$ range term in their interaction Lagrangian as these terms contain the number densities and scalar densities of hyperons respectively and not those of nucleons. Hence the observed net decrease of the in-medium mass in nuclear matter with respect to baryon density is mainly due to the $d_1$ range term which depends on the scalar densities of the nucleons. For $\eta=0.5$ and  $f_s=0.0$,  $d_1$ range term depends on the scalar densities of only neutrons. The scalar meson exchange term proportional to the fluctuation $\zeta^\prime$ in the medium contributes but their attractive contribution gets saturated at large densities owing to the saturation behaviour of $\zeta$ in the nuclear matter as a function of baryon density as shown in the figure \ref{sigma_zeta}. The magnitude of the repulsive first range term  which is also proportional to the fluctuation $\zeta^\prime$ gets surpassed by the dominant $d_1$ range term with  an increase in the density. Moreover, due to the absence of Weinberg-Tomozawa term, the self energy expression given by eq.(\ref{selfenergy_Ds}) becomes identical for both ${D_s}^+$ and ${D_s}^-$ mesons and  hence their in-medium masses are degenerate in the nuclear matter at a constant value of the magnetic field. In the nuclear medium, for isospin asymmetry parameter $\eta=0.5$ and at magnetic field  eB =$\ 4m_\pi^2$ incorporating the effects of AMM of neutrons, the in-medium masses of ${D_s}$ mesons are 1968.70 MeV, 1943.03 MeV, 1921.98 MeV and 1905.80 MeV at 1$\rho_0$, 2$\rho_0$, 3$\rho_0$ and $4\rho_0$ respectively.

In the strange hadronic medium, the contribution from Weinberg-Tomozawa term and $d_2$ range term is present in the interaction Lagrangian of $D_s$ mesons. The Weinberg-Tomozawa term which is attractive for ${D_s}^+$ meson and repulsive for ${D_s}^-$ meson, breaks their mass degeneracy in the medium when hyperons are present. This breaking of mass degeneracy gets aggravated with increase in baryon density. The variation of $\zeta$ is more pronounced in the hyperonic matter with respect to baryon density and hence the magnitude of contribution from scalar meson exchange term and first range term is larger compared to the nuclear medium.The net effect is in such a way that the overall attractive contributions from the total range term and scalar meson exchange term are dominant than the Weinberg-Tomozawa term which results in the mass drop of strange-charmed mesons in the magnetized hyperonic medium. In the case of ${D_s}^+$ meson the Weinberg-Tomozawa term aids the mass drop where as for ${D_s}^-$, the repulsive Weinberg-Tomozawa term decreases the overall attractive contributions from the other terms.
Hence the mass of ${D_s}^+$ meson drops more than ${D_s}^-$ mesons in the magnetized strange hadronic medium. In magnetized strange hadronic medium,  for isospin asymmetry parameter $\eta=0.5$, strangeness fraction $f_s=0.3$ and at magnetic field  eB =$\ 4m_\pi^2$ incorporating the effects of AMM of baryons, the in-medium masses of ${D_s}^+$ mesons are 1966.02 MeV, 1936.72 MeV, 1909.66  MeV and 1886.81 MeV
and the in-medium masses of ${D_s}^-$ mesons are 1969.16 MeV, 1942.92 MeV, 1918.83  MeV and 1898.98 MeV at 1$\rho_0$, 2$\rho_0$, 3$\rho_0$ and $4\rho_0$ respectively. 

It can be observed that at a fixed baryon density and magnetic field,  ${D_s}^+$ meson experiences a larger mass drop in strange hadronic matter compared to the nuclear matter at all densities due to the appearance of  Weinberg-Tomozawa term and $d_2$ range term (absent in nuclear matter) which are both attractive in nature. The increase in the magnitude of attractive scalar meson exchange term in hyperonic matter further drops the mass of  ${D_s}^+$ meson. In the case of ${D_s}^-$ meson, it experiences a lesser mass drop in the hyperonic matter compared to the nuclear matter at low to intermediate densities (up to 1.9$\rho_0$-2$\rho_0$). This is due to the appearance of repulsive Weinberg-Tomozawa term in hyperonic matter which becomes slightly dominant at this density regime. But at higher densities, ${D_s}^-$ drops more in hyperonic matter since the magnitude of the attractive contributions from $d_1$, $d_2$ range terms, and scalar meson exchange term becomes larger with increase in density and this dominates over the repulsive Weinberg-Tomozawa term. As $f_s$ increases, the mass drop becomes larger at large densities.

In the presence of the magnetic field, the in-medium masses of ${D_s}^+$ and ${D_s}^-$ mesons increases due to the Landau quantization effect similar to what we have observed in ${D}^\pm$ mesons. At small densities (up to $1.5\rho_0$) the increase in the in-medium mass of ${D_s}$ mesons due to this additional positive modification from the magnetic field will be even dominant over the mass drop due to the finite density contributions alone. In addition to this, the variation of the scalar field $\zeta$ and the scalar densities of baryons with respect to magnetic fields also contribute to the medium mass modification but is subdued by the positive contribution from the Landau quantization effect. As magnetic field is increased from eB =$\ 4m_\pi^2$ to eB =$\ 8m_\pi^2$, the in-medium mass of $D_s$ mesons increases by approximately 20 MeV and 28 MeV at $\rho_B$= $1\rho_0$ and $\rho_B$= $4\rho_0$ respectively when AMM effects are considered.  At a fixed value of the magnetic field, when AMM effects are not taken into account the in-medium masses of strange-charmed mesons are smaller in comparison to their masses when AMM effects of baryons are incorporated. This is due to the larger values of scalar densities and a smaller value of $\zeta$, resulting in  a slightly increased magnitude of the $d_1$, $d_2$ range terms as well as the scalar meson exchange term. The effects of AMM on the in-medium masses of these mesons becomes larger at larger magnetic fields and at larger densities. It can be observed from figures \ref{mD}, \ref{mDbar} and \ref{mD_s}, that the variation of the in-medium masses of ${D_s}$ mesons with respect to the increase in baryon density is less when compared to the variation of that of $D$ mesons. Hence the presence of magnetic fields play a relatively much more significant role in the overall medium mass modifications of ${D_s}$ mesons.

\begin{figure}[htbp]
\includegraphics[height=17.6cm, width=16.0cm]{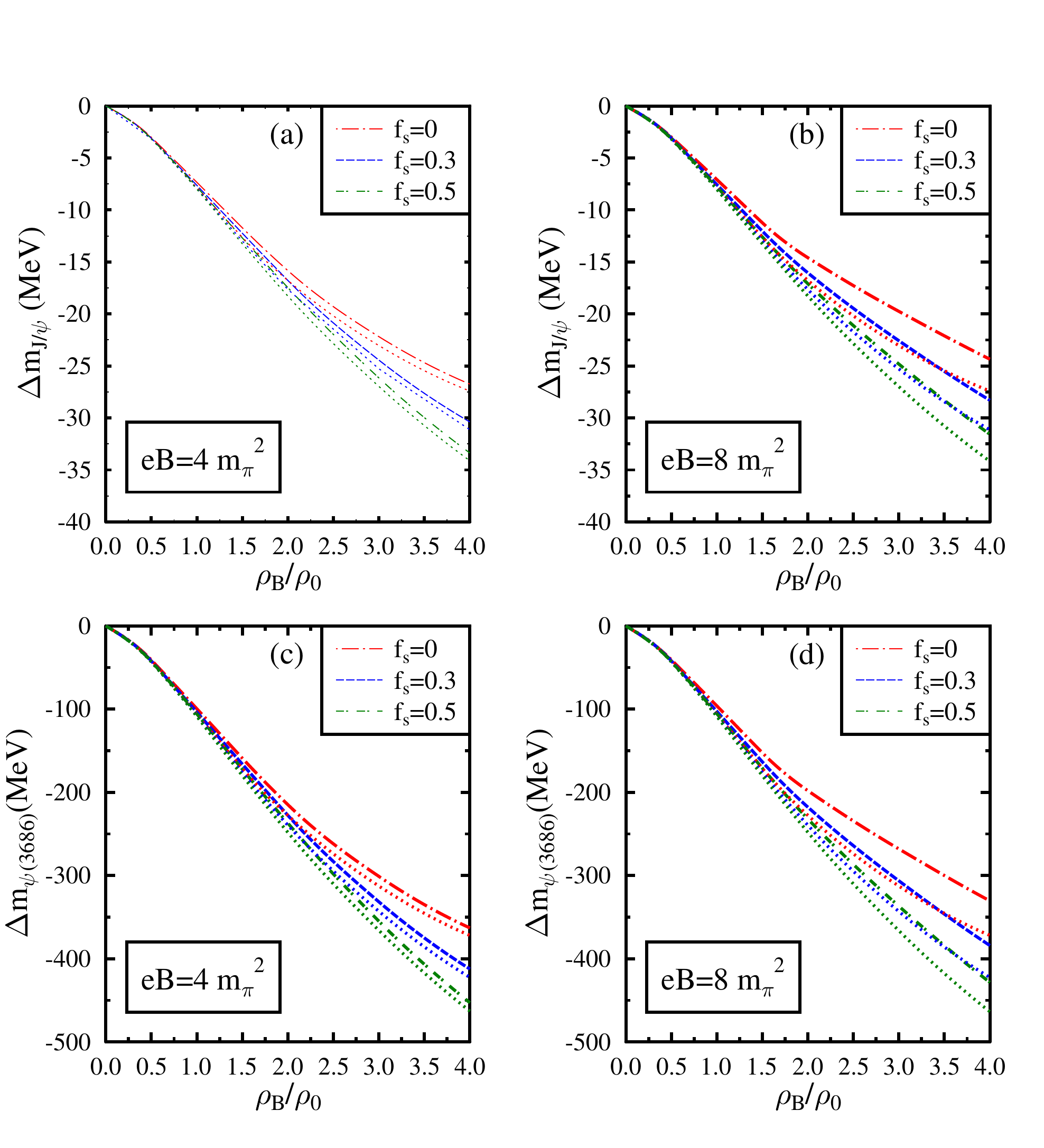}
\caption{The mass shift of $J/\psi$ and  $\psi(3686)$, in MeV, plotted as function of baryon density for both nuclear ($f_s$ = 0), and hyperonic ($f_s$ = 0.3, 0.5) matter situations at magnetic fields $eB=4m_{\pi}^2$ as well as for $eB=8m_{\pi}^2$, for isospin asymmetry parameter ($\eta=0.5$) , when the effects of anomalous magnetic moment are taken into account (dashed lines), and compared to the case when the effects of anomalous magnetic moment are not taken into account (dotted line)}
\label{jpsi_psi(3686)}
\end{figure}

\begin{figure}[htbp]
\includegraphics[height=17.6cm, width=16.0cm]{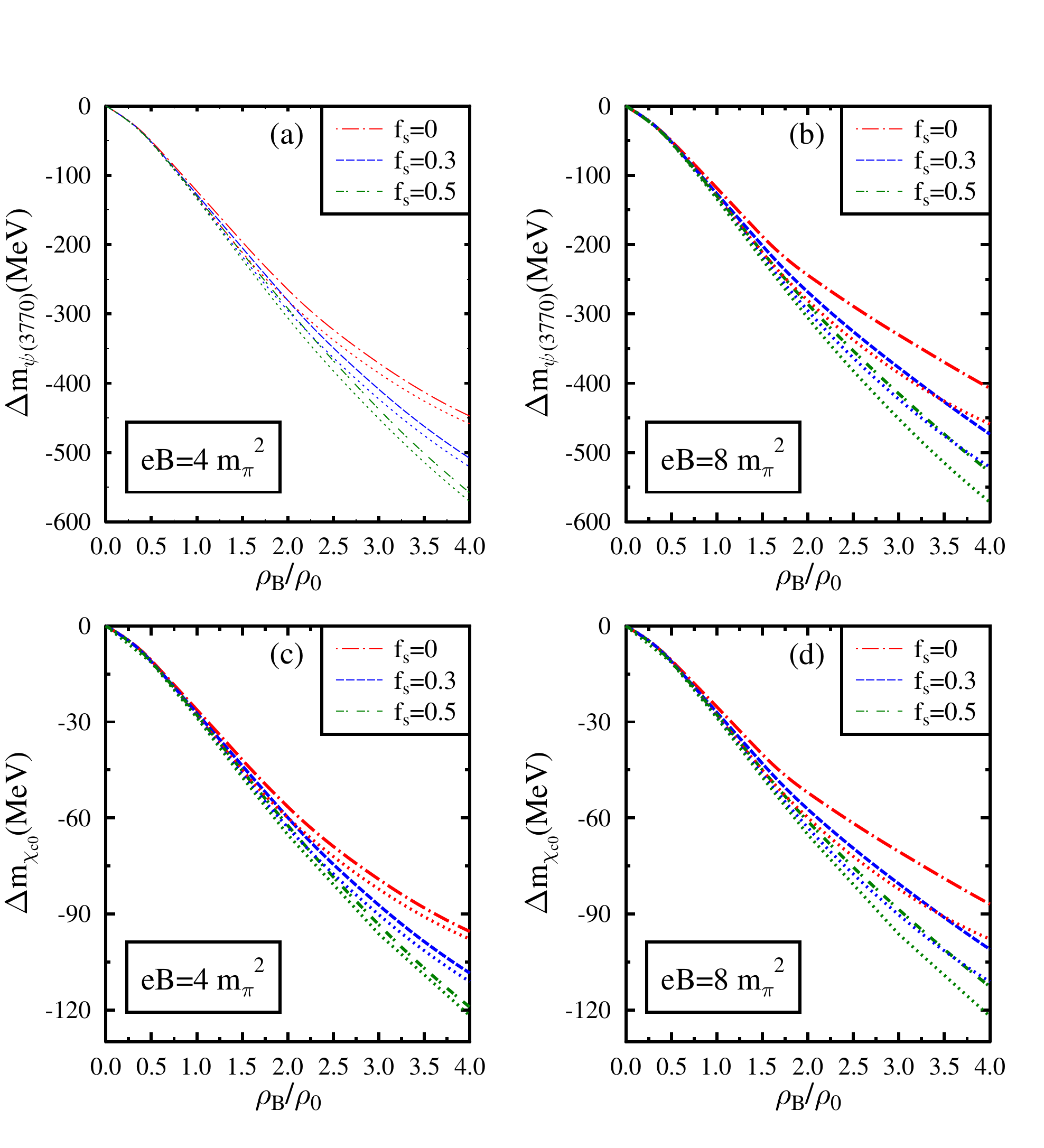}
\caption{The mass shift of  $\psi(3770)$  and $\chi_{c0}$ in MeV, plotted as function of baryon density for both nuclear ($f_s$ = 0), and hyperonic ($f_s$ = 0.3, 0.5) matter situations at magnetic fields $eB=4m_{\pi}^2$ as well as for $eB=8m_{\pi}^2$, for isospin asymmetry parameter ($\eta=0.5$) , when the effects of anomalous magnetic moment are taken into account (dashed lines), and compared to the case when the effects of anomalous magnetic moment are not taken into account (dotted line)}
\label{psi(3770)_chic0}
\end{figure}

\begin{figure}[htbp]
\includegraphics[height=8.8cm, width=16.0cm]{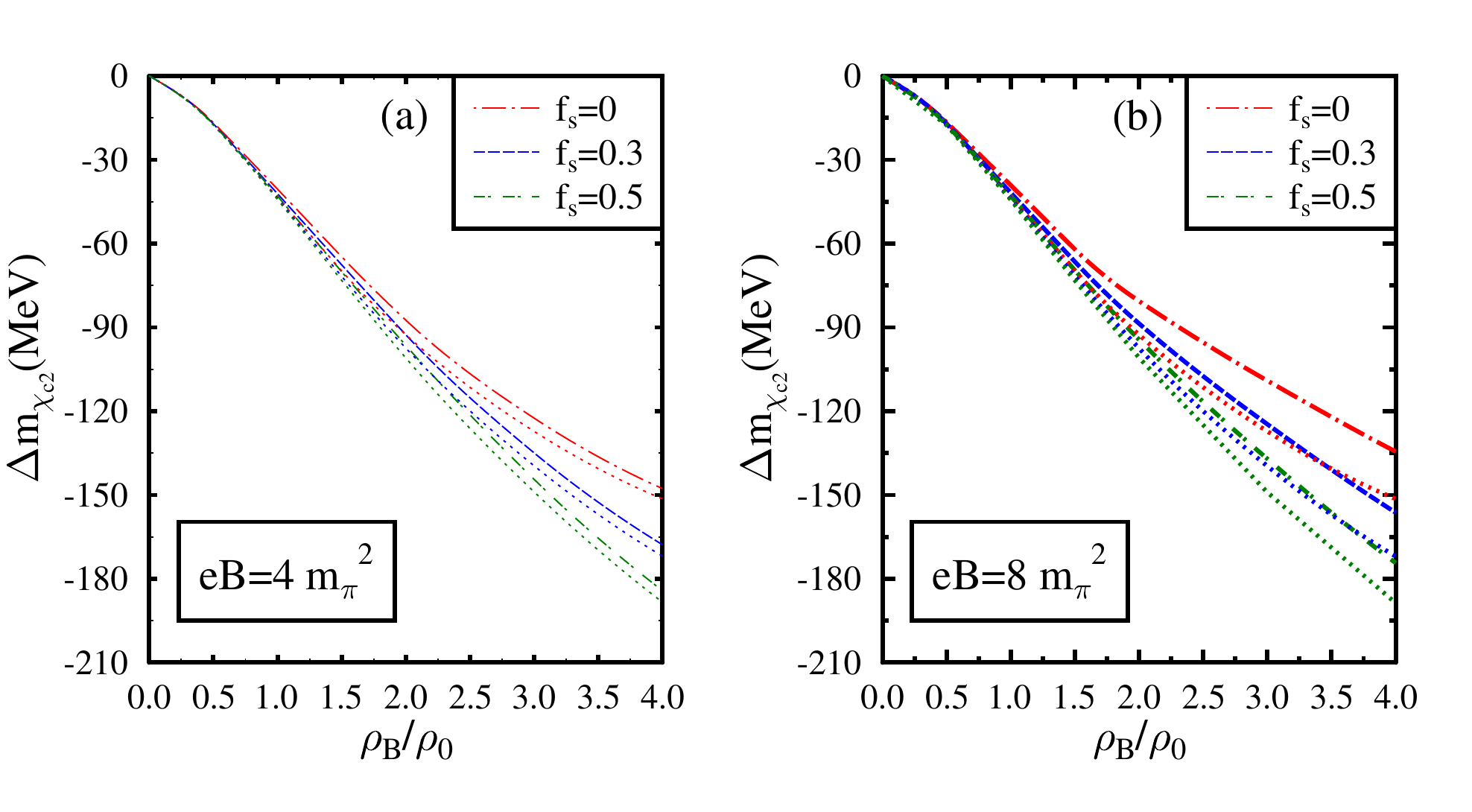}
\caption{The mass shift of $\chi_{c2}$ in MeV, plotted as function of baryon density for both nuclear ($f_s$ = 0), and hyperonic ($f_s$ = 0.3, 0.5) matter situations at magnetic fields $eB=4m_{\pi}^2$ as well as for $eB=8m_{\pi}^2$, for isospin asymmetry parameter ($\eta=0.5$) , when the effects of anomalous magnetic moment are taken into account (dashed lines), and compared to the case when the effects of anomalous magnetic moment are not taken into account (dotted line)}
\label{chic2}
\end{figure}

The medium modifications of the charmonium states in the strange hadronic matter under strong magnetic fields are calculated from the medium change of the dilaton field $\chi$. The mass shifts of charmonium states $J/\psi$, $\psi(3686)$, $\psi(3770)$, $\chi_{c0}$ and $\chi_{c2}$ from their vacuum masses are plotted as a function of nuclear densities at magnetic fields $\ 4m_\pi^2$ and $\ 8m_\pi^2$ for $\eta=0.5$ and for various values of $f_s$ in figures \ref{jpsi_psi(3686)}, \ref{psi(3770)_chic0} and, \ref{chic2}. The effects of anomalous magnetic moment are also incorporated into these plots represented by dashed lines. The value of the mass of the charm quark is taken to be $\ m_c=1.95$ GeV in the present investigation. Such a choice could reproduce the mass difference of the charmonium states $J/\psi$ and $\psi(3686)$ in vacuum\cite{CharmoniumLee}. The charmonium system can be described as a non-relativistic bound state of massive charm quark and an anti-quark that interacts via an inter-quark potential. In this investigation the wave functions of states are  taken to be harmonic oscillator wave functions. The value of the parameter $\beta$ which characterizes the strength of the harmonic potential, in GeV for $J/\psi$, $\psi(3686)$ and  $\psi(3770)$  are 0.51, 0.38 and 0.37. They are calculated using their mean square radii of these charmonium states  which are  $0.47^2$$fm^2$, $0.96^2$$ fm^2$ and $1 fm^2$ respectively\cite{CharmoniumLee}. For the $\chi_{c0}$ and $\chi_{c2}$ states, the values of $\beta$ in GeV are taken to be 0.44  and 0.41 by linear extrapolation of the vacuum mass versus $\beta$ graph of the charmonium states $J/\psi$, $\psi(3686)$ and  $\psi(3770)$\cite{AM_charmdecaywidths_mag}.

The dominant medium effect on the mass modifications of the charmonium states in the hadronic medium is observed to be the effect of density, which is evident from the plots. At a particular magnetic field and strangeness fraction, as baryon density increases, there is a drop in the value of $\chi$ from its vacuum value. Hence the mass shifts of all charmonium states (proportional to $\chi^4-{\chi_0}^4$) steadily increase. The magnitudes of the mass shifts for each charmonium state are proportional to the magnitude of the integral (eq.(\ref{masspsi}) calculated from their respective momentum wave functions. The mass drop of $J/\psi$  turns out to be small compared to other excited charmonium states which have significant mass drops in the magnetized strange hadronic medium. The mass modifications of $\psi(3770)$ as a function of the baryon density is the largest among all the charmonium states. This is because the momentum dependant integral calculated for this particular state amplifies the density dependence of the mass shift. The magnitude of the mass drop increases for the higher excited states of charmonia for a fixed value of density. Since the value of dilaton field $\chi$ decreases with increase in strangeness fraction of the medium, the mass modifications of charmonia are larger in the hyperonic medium compared to the nuclear medium.

For a fixed value of baryon density and strangeness fraction, when we account the effects of the anomalous magnetic moment, as the magnetic field increases, the mass modifications of charmonium states mostly remain constant with negligible variation at lower densities (up to 2 $\rho_0$ ). But at higher densities, the mass drop of charmonium states decreases as the magnetic field becomes larger because of the increase in the value of $\chi$ with the magnetic field. 
 As density increases the effect of the magnetic field is more prominent. Moreover, the mass shift from the magnetic field is more in nuclear matter compared to hyperonic matter. For the asymmetric hyperonic matter, without AMM, the mass shift is larger compared to the case where AMM effects are incorporated and this difference gets magnified as the strength of the magnetic field increases. Owing to the negligible modification of dilaton field $\chi$, the mass modifications of charmonium states at a  fixed density remains mostly independent with respect to change in the  magnetic field when AMM effects of baryons are neglected.
 
For isospin asymmetry parameter $\eta=0.5$ and at strangeness fraction $f_s$=0.3, under a magnetic field of eB= $\ 4m_\pi^2$ incorporating the effects of AMM, the mass shifts (in MeV) of $J/\psi$, $\psi(3686)$, $\psi(3770)$, $\chi_{c0}$, $\chi_{c2}$ are $-7.63$ $(-30.38)$, $-103.59$ $(-412.36)$, 
$-127.68$ $(-508.26)$,$-27.25$ $(-108.48)$ and $-42.15$ $(-167.80)$ respectively at $\rho_B$= 1$\rho_0$(4$\rho_0$), where as at eB= $\ 8m_\pi^2$  the mass shifts of charmonia under similar medium conditions in the same order are $-7.57$ $(-28.30)$, $-102.76$ $(-384.11)$, $-126.66$ $(-473.45)$, $-27.03$ $(-101.05)$ and  $-41.81$ $(-156.30)$  respectively. When the effects of AMM of baryons are ignored, for $\eta=0.5$ and at $f_s$=0.3, under a magnetic field of eB= $\ 4m_\pi^2$, the mass shifts are observed to be $-7.94$ $(-31.12)$, $-107.81$ $(-422.43)$, $-132.89$ $(-520.67)$, $-28.36$ $(-111.13)$ and $-43.87$ $(-171.89)$ at $\rho_B$= 1$\rho_0$(4$\rho_0$). In the absence of AMM , when the magnetic field is increased to eB= $\ 8m_\pi^2$,the mass shifts of charmonia become $-7.95$ $(-31.13)$, $-107.93$ $(-422.57)$, $-133.03$ $(-520.84)$, $-28.39$ $(-111.17)$ and $-43.92$ $(-171.95)$.

When the strangeness fraction is increased to $f_s$=0.5, the mass shifts (in MeV) of $J/\psi$, $\psi(3686)$, $\psi(3770)$, $\chi_{c0}$, $\chi_{c2}$ are $-7.79$ $(-33.36)$, $-105.73$ $(-452.84)$, $-130.32$ $(-558.16)$, $-27.81$ $(-119.14)$ and	$-43.02$ $(-184.27)$ respectively at $\rho_B$= 1$\rho_0$(4$\rho_0$) under a magnetic field of eB= $\ 4m_\pi^2$ with the effects of AMM included. At eB= $\ 8m_\pi^2$  the mass shifts under similar medium conditions are observed to be $-7.77$ ($-31.56$), $-105.48$($-428.42$), $-130.01$ ($-528.05$), $-27.75$($-112.71$) and $-42.92$($-174.33$) respectively. When AMM effects are ignored, under a magnetic field of eB= $\ 4m_\pi^2$, the mass shifts of these charmonium states are observed to be $-8.00$ $(-34.11)$, $-108.68$ $(-462.94)$, $-133.95$ $(-570.61)$, $-28.59$ $(-121.79)$ and $-44.22$ $(-188.38)$ at $\rho_B$= 1$\rho_0 $(4$\rho_0$) and these values modify negligibly when the magnetic field is increased to eB= $\ 8m_\pi^2$ as seen before in the $f_s$=0.3 case.

In Ref \cite{CharmoniumLee}, the in-medium masses of $J/\psi$, $\psi(3686)$ and $\psi(3770)$ in the nuclear matter without taking magnetic field into account are calculated to be $-8$, $-100$ and $-140$ MeV  at $\rho_B$= $\rho_0$ using linear density approximation from the QCD second order stark effect. The mass-shift for $J/\psi$ has also been investigated with in the QCD sum rules in Ref\cite{Klingl} without taking the magnetic field into account and the value of the mass shift at nuclear saturation density was observed to be about $-7$ MeV. In the absence of the magnetic field, the chiral effective model predicts a mass shift of $-8.6$, $-117$, $-155$  MeV in  symmetric ($\eta=0 $) nuclear matter and $-8.4$, $-114$, $-150$ MeV in asymmetric($\eta=0.5$) nuclear matter for $J/\psi$, $\psi(3686)$ and $\psi(3770)$ \cite{AKumar_AM_EurPhys2011}. In the symmetric hyperonic medium ($f_s=0.5$) the mass shifts of the above charmonium states calculated within the chiral model, without considering the magnetic field are $-8.41$, $-114$ and, $-151$ MeV, taking $N_f$ = 3 in the beta function, to obtain the expression for the gluon condensates\cite{AKumar_AM_EurPhys2011}. In Ref \cite{Amal1}, the numerical results of charmonia masses in the magnetized nuclear matter are given in detail for the symmetric case as well as in the asymmetric case.

\section{SUMMARY}
The medium mass modifications of open charm mesons ($D$, $\bar{D}$, $D_{s}$), and charmonium states in the strange hadronic matter in the presence of strong magnetic fields are investigated using a chiral effective Lagrangian model. In high energy heavy Ion Collision experiments where strong magnetic fields are created, the study of mass modifications of hadrons in the presence of magnetic fields is necessary, as they affect the experimental observables. The magnetic field will distinguish charged baryons from neutral baryons in the medium due to differences in their charge and anomalous magnetic moments. The number density and scalar density of charged baryons have contributions from Landau energy levels. We have studied the isospin asymmetric magnetized strange hadronic matter by solving the equations of motion to obtain the values of scalar fields as functions of density and strangeness fraction at various magnetic fields (including effects of anomalous magnetic moments of baryons). When AMM effects are incorporated, at high densities the scalar fields $\sigma$, $\zeta$, and $\chi$ are subjected to a decrease in their magnitude as baryon density and strangeness fraction increases whereas $\delta$ exhibits a saturation behaviour under similar conditions. The obtained values of scalar fields are used to find the mass modifications of open charm mesons and charmonia. In general, the masses of all these mesons decrease with an increase in density. The $D$, $\bar{D}$ ,${D_{s}}^+$ mesons, and charmonium states experience a larger mass drop in the magnetized hyperonic matter compared to the nuclear matter. But ${D_{s}}^-$ exhibits a counter behaviour at low densities in the magnetized strange hadronic medium. The mass shifts of charmonia with respect to the magnetic field are negligible when AMM effects are neglected and the excited states experience a larger mass shift. The charged ${D}^+$, ${D}^-$,${D_{s}}^+$ and ${D_{s}}^-$ mesons have additional positive mass shifts due to Landau quantization effects in the presence of the magnetic field. This effect becomes much more significant for ${D_{s}}$ mesons, especially at low densities. In the present investigation of study of open charm mesons and charmonium states, the dominant medium effect is due to the density as compared to the magnetic field which should have observable consequences in the ${D}^+$/${D}^0$, ${D}^-$/ $\bar{D^0}$ and ${D_{s}}^+$/ ${D_{s}}^-$ ratios in asymmetric heavy ion collisions in Compressed Baryonic Matter (CBM) experiments at FAIR at the future facility of GSI.

\section{ACKNOWLEDGEMENTS}
 A.J.C.S acknowledges the support towards this work from the Department of Science and Technology, Government of India, via an INSPIRE fellowship (INSPIRE Code IF170745). A.J.C.S is thankful to Ankit Kumar, Manju Soni, Jaswant Singh and, Arjun Kumar for fruitful discussions. AM acknowledges financial support from Department of Science and Technology (DST), Government of India (project no.CRG/2018/002226).

\end{document}